\documentclass[aps,prx,twocolumn,showpacs,groupedaddress,reprint,longbibliography]{revtex4-1}

\usepackage{tabularx}
\usepackage{longtable}
\usepackage{epsfig}
\usepackage{epstopdf}

\newcommand{\om}{\omega}

\newcommand{\be}{\begin{equation}}
\newcommand{\ee}{\end{equation}}
\newcommand{\bea}{\begin{eqnarray}}
\newcommand{\eea}{\end{eqnarray}}

\newcommand{\bo}{\beta_\oplus}
\newcommand{\cX}{c^{\rm T}_X}
\newcommand{\cY}{c^{\rm T}_Y}
\newcommand{\cZ}{c^{\rm T}_Z}
\newcommand{\cTX}{c^{\rm T}_{TX}}
\newcommand{\cTY}{c^{\rm T}_{TY}}
\newcommand{\cTZ}{c^{\rm T}_{TZ}}

\begin{document}

\title{Acoustic tests of Lorentz symmetry using quartz oscillators
%Testing the isotropy of space using rotating quartz oscillators
}

\author{Anthony Lo}
\affiliation{Department of Physics, University of California, Berkeley, CA 94720, USA}
\author{Philipp Haslinger}
\author{Eli Mizrachi}
\author{Lo\"ic Anderegg}
\author{Holger M\"{u}ller}
\email{hm@berkeley.edu}
\affiliation{Department of Physics, University of California, Berkeley, CA 94720, USA}
\author{Michael Hohensee}
\altaffiliation{Department of Physics, University of California, Berkeley, CA 94720, USA}
\affiliation{Lawrence Livermore National Laboratory, Livermore, CA 94550, USA}
\author{Maxim Goryachev}
\author{Michael E Tobar}
\affiliation{ARC Centre of Excellence for Engineered Quantum Systems, School of Physics, University of Western Australia, 35 Stirling Highway, Crawley WA 6009, Australia}

\date{\today}

%List of changes:
%Emphasize LI of the photon in abstract and somewhere in text
%Emphasize that all coefficients need to be measured to have solid case

\begin{abstract}
We propose and demonstrate a test of Lorentz symmetry based on new, compact, and reliable quartz oscillator technology. Violations of Lorentz invariance in the matter and photon-sector of the standard model extension (SME) generate anisotropies in particles' inertial masses and the elastic constants of solids, giving rise to measurable anisotropies in the resonance frequencies of acoustic modes in solids. A first realization of such a ``phonon-sector" test of Lorentz symmetry using room-temperature SC-cut crystals yields 120\,hours of data at a frequency resolution of $2.4\times 10^{-15}$ and a limit of $\tilde c_Q^{\rm n}=(-1.8 \pm 2.2)\times 10^{-14}$\,GeV on the most weakly constrained neutron-sector $c-$coefficient of the SME. Future experiments with cryogenic oscillators promise significant improvements in accuracy, opening up the potential for improved limits on Lorentz violation in the neutron, proton, electron and photon sector.
\end{abstract}
\widetext

%\keywords{}

\maketitle

%\tableofcontents

\draft
\preprint{LLNL-JRNL-664409-DRAFT}

\section{Introduction}

%Experimental tests of Lorentz symmetry go back at least to A. A. Michelson's 1881 search for anisotropy in the propagation speed of light.
The possibility that physics beyond the standard model might violate Lorentz invariance \cite{ColladayKostelecky,KosteleckyMewes,KosteleckyMewesPRD} has motivated experimental tests with high precision and broad scope. In particular, experiments have placed stringent limits on anisotropies in the laws of motion of the photon, electron, proton and neutron based on, e.g., electromagnetic cavities \cite{Achim,Stanwix,Schiller,TLVMC}, clock comparisons \cite{KosteleckyLane,WolfPRD2006,Flambaum2009,Crane,Bear,Dy}, magnetometry \cite{Peck,Brown,Gemmel,Altschul2009,Smiciklas}, ultracold neutrons \cite{Altarev} and ion traps \cite{Haeffner}. Some anisotropic inertial masses of particles are known to be below $10^{-28}$\,GeV \cite{Smiciklas}, but others are more weakly constrained. Bounding all modes of Lorentz violation often requires active rotation and Earth's orbit to modulate the orientation and velocity of the apparatus, and thus data taking over a year \cite{Crane,Heckel,Bluhm,Gomes}, but at the same time uses fragile and maintenance-intensive atomic and optical setups. In this paper, we introduce the concept of an acoustic test of Lorentz symmetry, based on precision measurements of phonon oscillations in a quartz crystal oscillator, demonstrating a simple and reliable, yet sensitive method that is readily suited for long-term operation on a turntable or even being carried on small air and space vehicles. Based on commercial SC-cut quartz oscillators, we limit the most weakly constrained mode of neutron-sector violations in the SME to $(-1.8\pm 2.2)\times 10^{-14}$\,GeV, improving on previous laboratory experiments~\cite{KosteleckyTasson} by three orders of magnitude and on a previous astrophysics bound by about one. This rules out all possibilities for Lorentz-violating anisotropies in the inertial mass of neutrons, protons and electron at the $\sim 10^{-14}$\,GeV level. We show that future experiments with cryogenic oscillators could be used to perform more sensitive tests of Lorentz symmetry in the proton, neutron, electron, and photon sectors.

Our method is the first to compare acoustic oscillations in different directions to constrain Lorentz symmetry. In this work we show that the frequencies of the bulk elastic waves  are sensitive to the photon-, electron-, proton- and neutron-sector of SME. The coefficients of the SME that change the resonance frequency of acoustic modes in solids will be collectively referred to as phonon-sector coefficients, in analogy to the photon-sector.  This work also paves the way for future experiments that involve high-Q frequency stable phonon systems, which could be adapted to test Lorentz symmetry. Such systems include but are not limited to, phonon lasers \cite{phononlaser}, optomechanical systems \cite{luan,seok,VahalaAPL} and phonon induced Brillouin scattering devices \cite{VahalaScience,Tomes,Grudinin,Lee,Li}.%It is based on compact, reliable technology This allows further experiments where changing gravitational potentials are desirable, such as studies of the equivalence principle.
%Our oscillators are relatively insensitive to external influences such as temperature changes and vibrations.  %However, they are sensitive to many sectors of Lorentz symmetry violation (e.g. photon, neutron, proton and electron) because the properties of all these affect the speed of sound in solids.

Hundreds of limits on Lorentz invariance violations do exist \cite{datatables}, but there are as many gaps where large signals may lie undetected. The gaps are typically left behind by other technologies because filling them would take rotating setups, long-term data taking, or operation in difficult environments. Our method is well suited for this task. We do note, however, that nonstandard phonon-sector signals may exist for reasons other than encoded in the SME, and this may lead to genuinely new tests of fundamental laws of physics. %Examples may include searches for topological dark matter by a worldwide clock network \cite{PospelovDerevianko}.

\section{Theory}
\subsection{Standard Model Extension}
Lorentz invariance violation has been parameterized in several ways, e.g., \cite{Lightman1973,Nielsen1983,Will1993,Coleman1999}.
We use a phenomenological framework known as the Standard Model Extension (SME) \cite{ColladayKostelecky,KosteleckyMewes,KosteleckyMewesPRD} to describe the effects of Lorentz violation. It augments the Standard Model with new combinations of known particles and fields that lead to Lorentz violation, subject to the requirement that the theory must respect conservation of energy and momentum, renormalizability, gauge-invariance, and observer Lorentz covariance. % (i.e., an experimental outcome may depend on its components� motion, but not on the coordinates used to describe it).
The new terms in the SME are parameterized by tensors, whose component values are collectively known as Lorentz-violation coefficients. If all such coefficients are zero, Lorentz symmetry is exact.
The value of these Lorentz-violating coefficients are by definition frame-dependent, but can be taken as approximately constant in any frame that is inertial on all time-scales relevant to the experiment. It is conventional to use a sun-centered celestial equatorial reference frame. Quantities in this frame will be denoted by capital indices $T,X,Y,Z$. The time coordinate $T$ has its origin at the 2000 vernal equinox. The $Z$ axis is directed north and parallel to the rotational axis of the Earth at $T =0$. The $X$ axis points from the Sun towards the vernal equinox, while the $Y$ axis completes a right-handed system \cite{datatables}.

%The model makes no prediction for the size of the coefficients, which thus need to be measured experimentally. There are, however,

\subsection{The $c$ coefficients}
We will study the influence of $c$-type coefficients in detail here. Later, we will give an overview of all Fermion- and photon-sector coefficients of the minimal SME and what levels of sensitivity can be expected for them in phonon-sector experiments.

The coefficients $c_{\mu\nu}$ enter the Lagrangian of a free Dirac fermion $w$ by substituting the Dirac matrix $\gamma \rightarrow \gamma_\nu + c_{\mu\nu}\gamma^\mu$, where $\nu=0,1,2,3$ are the space-time coordinates \cite{KosteleckyLane}. The term enters the  non-relativistic Schr\"odinger hamiltonian of a particle by the substitution
\be\label{h}
\frac{p^2}{2m}\rightarrow \frac{p_j p_k}{2m} (\delta_{jk}-2c^{w}_{jk}-c_{00}\delta_{jk}),
\ee
%or alternatively, in the non relativistic Lagrangian,
%\be
%\frac 12 m v^2 \rightarrow \frac 12 m v^j v^k (\delta_{jk}+2c^{w}_{jk}+c_{00}\delta_{jk}),
%\ee
where $p_j$ %and $v^j$
are the components of momentum, $j,k=1,2,3$, and $m$ is the particle mass. Thus, the $c$ coefficients describe anisotropies of the inertial mass of particles that depend on the direction of its motion.  (Though protons and neutrons are composite particles, they are approximated as Dirac fermions for the purpose of parameterizing Lorentz violation at low energies.)  The electron, proton, and neutron tensors $c^{\rm e}_{\mu\nu}$, $c^{\rm p}_{\mu\nu}$, and $c^{\rm n}_{\mu\nu}$, are independent of one another, symmetric, and traceless with 9 independent degrees of freedom each.  %The opposite signs of the $c$'s in the two expressions are consistent with the definition $p_k=\partial L/(\partial \dot x^k)$.)
%Each tensor $c^{w}_{\mu\nu}$ is symmetric and traceless and thus has 10 independent degrees of freedom.

For a composite object T that consists of $n^{\rm w}$ particles of species w, the effects of Lorentz violation in Eq. (\ref{h}) are given by effective coefficients
\be\label{cavg}
c^{\rm T}_{\mu\nu}=\frac{1}{m^{\rm T}}\sum_{\rm w} n^{\rm w} m^{\rm w} c^{\rm w}_{\mu\nu}, \quad m^{\rm T} = \sum_{\rm w} n^{\rm w} m^{\rm w}.
\ee
Not all of the basic $c_{\mu\nu}$ coefficients are physical. The physical combinations are conventionally expressed by the combinations \cite{datatables}
\begin{eqnarray}\label{cs}
\tilde c_Q^{w}&=&m^w(c_{XX}^w+c_{YY}^w-2c_{ZZ}^w), \nonumber \\
\tilde c_{-}^w&=&m^w(c_{XX}-c_{YY}), \nonumber \\
\tilde c_{J}^w&=&m^w|\varepsilon_{JKL}|c_{KL}, \nonumber \\
\tilde c_{TJ}^w&=& m^w(c_{TJ}+c_{JT}),\nonumber \\
\tilde c_{TT}^w &=&m^wc_{TT}.
\end{eqnarray}
We will use these combinations throughout when stating experimental results. We note that by these definitions, $\tilde c_{TJ}\neq mc_{TJ}$.

\subsection{Relative significance of the terms}

The components of $\tilde c$ encode independent degrees of freedom for Lorentz violation. Knowledge of one or many of them does not imply anything about the remaining ones. By analogy, in the photon sector several different modes of Lorentz violation exist, that are characterized by how they transform under Lorentz boosts and rotations. Some of them, the $\tilde \kappa_{e+}$ and $\tilde \kappa_{o-}$ have been bounded astrophysically to an accuracy of $10^{-34}$ \cite{Mewes}. Other coefficients (denoted $\tilde \kappa_{e-}$ and $\tilde \kappa_{o+}$) remain unexplored by these type of observations. They are best studied by laboratory experiments, which have been continuously improved from the original ones by Michelson-Morley, Kennedy-Thorndike and Ives-Stilwell \cite{MM1,*MM2,*MM3,*MM4} to modern ones that reach down to sensitivities of $10^{-18}$ \cite{Photons1,*Photons2,*Photons3,*Photons4,*Photons5,*Photons6,*Photons7,*Photons8,*Photons9,*Photons10,*Photons11,*Photons12,*Photons13,Schiller,Achim,Stanwix,Nagel}. But even those tests leave behind a last remaining coefficient. This one, $\tilde \kappa_{\rm tr}$,  is arguably the hardest to measure, as dedicated experiments were set up to measure this remaining coefficient, and now all modes of Lorentz violation in the minimal photon sector have been very stringently limited by experiment\cite{Stanwix,Hohensee,Baynes}.

By comparison, the Fermion sectors have been studied less comprehensively. Here also, some components of the coefficients for Lorentz violation have been limited with high precision, but others remain tested at low precision.

\subsection{Influence of the Fermion $c$-coefficients in crystal oscillators}

The principle of our search for anisotropic inertial masses is simple: We use a quartz oscillator performing nominally 10-MHz oscillations on a turntable. If the inertial mass of the quartz material in on direction is fractionally higher by $\delta m/m$ than in an orthogonal direction, then rotating the crystal leads to a modulation of the oscillation frequency by $\frac{\delta\nu}{\nu}=-\frac 12 \delta m/m$. We use shear oscillations in an sc-cut quartz crystal. This modulation can be measured, either by comparison to a stationary reference or by comparison to a second oscillator on the turntable, rotated by $90^\circ$ relative to the first.

Finding the sensitivity of mechanical resonators to Lorentz violation is possible by perturbation theory for each eigenmode. Our experiment uses a Stress Compensated (SC) cut \cite{pz:1988zr} crystal Bulk Acoustic Wave (BAW) piezoelectric plate resonator working at the third overtone (OT) of the thickness shear mode. This resonator is housed in an oven at the temperature of around 85$^\circ$C where the vibrational mode exhibits zero temperature coefficient of its oscillation frequency. The SC cut is doubly rotated relative to the crystal axis by a first angle of $\theta=34.11^\circ$ and a second angle $\phi=21.93^\circ$. This also results in zero stress dependence of the frequency, which reduces dependence of the frequency on the mounting of the crystal, amplitude variations of the oscillation, and ageing \cite{Filler:1988oa}.

\subsubsection{Unperturbed modes}

The eigenmodes of doubly rotated piezoelectric plate resonators have been studied in detail \cite{Stevens,Eernisse}. Due to high $Q$-factors (typically slightly above $10^6$ at room temperature) the eigenmodes may be considered isolated mechanical oscillators. We introduce a plate coordinate system $x^{[i]}$ ($i=1,2,3$) in which $x^{[2]}$ is normal to the major surfaces of the doubly rotated quartz blank, $x^{[1]}$ is directed along the axis of the second rotation, and $x^{[3]}$ is completing a right-handed system (Tab. \ref{coordinates}). %In our experiment, $x^{[2]}$ is oriented vertically and $x^{[1]}$ horizontally.
We denote $u^{[i]}(t, x)$ the components of the displacement of a volume element at $x^{[i]}$ as function of time $t$. We start by finding the modes that depend only on the $x^{[2]}$ coordinate (``thickness modes"),
\be
u^{[r]}=A^{[r]}\sin (\eta x^{[2]}) e^{i\omega t},
\ee
where
\be
(\bar{\hat{c}}_{{[2nr2]}}-\bar c \delta_{{[nr]}})A^{[r]}=0.
\ee
The $\bar c_{[2nr2]}$ are the piezoelectrically stiffened elastic constants rotated into the blank coordinate system. Solving the last equation yields three eigenvectors $(A^{(1-3)})^{[r]}$. These eigenvectors are used as the basis of new ``thickness" coordinates $x^{(1-3)}$, organized such that $x^{(i)}$ has its largest component along $x^{[i]}$. Analysis in the thickness mode coordinates then yields three mode families, known as quasi-longitudinal (A-) mode, fast shear (B-) mode and slow shear (C-) mode. For each family, the amplitude of one of the displacement components in $x^{(i)}$ direction is large while the others are small. Due to this smallness, the modes nearly decouple, which makes it possible to find accurate closed-form expressions for the eigenmodes.

The modes of interest here have the largest displacement component along $x^{(1)}$ which is approximately along the $x^{[1]}$ axis. %, which is oriented vertically in our setup because of the way the coordinates are constructed.
It can be written as \cite{Eernisse}
\be
u_{1nmp}=e^{-\alpha_{1n} x_1^2/2}H_m\left(\sqrt{\alpha_{1n}}x_1\right)e^{-\beta_{1n} x_3^2/2}H_p(\sqrt{\beta_{1n}} x_3)
\ee
where
\be
\alpha^2_{1n}=\frac{n^2\pi^2\hat c^{(1)}}{8Rh_0^3M'_{1n}},\quad \beta^2_{1n}=\frac{n^2\pi^2\hat c^{(1)}}{8Rh_0^3P'_{1n}}
\ee
and $H_m$ is the Hermite polynomial of order $m$. For the mode used in a third-overtone sc cut crystal at 10\,MHz, we have $n=3$, $m=p=0$, $M'_{1n}=5.3273$, $P'_{1n}=6.3858$, $\hat c^{(1)}=3.4379$, $R$ is the radius of the blank and $h_0$ is the thickness, which is 0.54094\,mm to make the resonance frequency 10\,MHz \cite{Eernisse}.

\begin{table}
\caption{\label{coordinates} Coordinates used in this paper}
\begin{tabular}{ccp{2in}}\hline\hline
Name & Notation & Description \\ \hline
Blank & $x^{[1]}$ & Axis of second crystal rotation; approximately direction of shear \\
  	& $x^{[2]}$ & Normal to major blank surface \\
  	& $x^{[3]}$ & Completes right-handed system \\
Thickness & $x^{(i)}$ & Parallel to thickness modes \\
Lab & $x^1=x$ & Horizontally pointing South \\
	& $x^2=y$ & Horizontally pointing East \\
	& $x^3=z$ & Vertically upwards \\
Sun-centered & $x^T=T$ & $T=0$ at 2000 vernal equinox \\
         	& $x^X$ & From Sun towards vernal equinox \\
         	& $x^Y$ & Completes right-handed system \\
         	& $x^Z$ & Parallel to Earth's axis pointing North \\ \hline\hline
\end{tabular}
\end{table}

\subsubsection{Perturbation due to Lorentz violation}

%The motion of the volume elements in our crystal oscillator is predominantly in the $x^{[1]}$ direction, and the influence of Lorentz violation is given by Eq. (\ref{h}). This is easy to evaluate.
Since the motion of the volume elements is primarily in $x^{[1]}$ direction, Eq. (\ref{h}) predicts that
\be
\frac{p_{[x]}^2}{2m}\rightarrow \frac{p_{[x]}^2}{2m}\left(1-2c_{[xx]}-c_{[00]}\right),
\ee
which is equivalent to a re-scaling of the inertial mass by $1+2c_{[xx]}+c_{00}$. Solids are composite objects; summing up the contributions of the re-scalings in the electron, proton, and neutron sectors amounts to replacing the coefficients by the effective coefficients Eq. (\ref{cavg}). This leads to a relative change in the resonance frequency of
\[
\frac{\delta \nu}{\nu}=-\frac 12 (2c^{\rm T}_{[xx]}+c^{\rm T}_{00}).
\]

To study a simple case first, we may assume that all coefficients of Eq. (\ref{cs}) were zero except for $\tilde c_Q$, that the experiment with two rotating quartz oscillators was located at the equator with the $[x]$-axis horizontal and rotated around a vertical axis at an angular velocity of $\om_t$. This would lead to a modulation amplitude of $\delta\nu/\nu =c^{\rm Q}_Q/4$, where the superscript Q indicates we are using the effective combination of coefficients for quartz.

For the general case, we calculate the components $c_{[xx]}$ in the crystal frame, rotating on the turntable, from the $c_{\mu\nu}$ in the sun-centered frame. This involves Lorentz boosts and rotations \cite{KosteleckyLane}. We denote $\omega_t$ the angular velocity of the turntable measured in the lab frame, $\omega_\oplus \approx 2\pi$/(23h56min), and $\Omega_\oplus = 2\pi$/(1\,year) the sidereal angular velocities of Earth's rotation and orbit, respectively; $\chi$ is the co-latitude of the lab in which the experiment is performed ($\chi \approx 52.13^\circ$ for the current experiment in Berkeley, California); and $\eta \approx  23.4^\circ$ the angle between the ecliptic and Earth's equatorial plane.

The signal includes contributions of order 1, or suppressed by either the Earth's orbital velocity $\beta_\oplus \approx 10^{-4}$. We neglect contributions from signals suppressed by higher powers of $\beta_\oplus$ or by the velocity of the laboratory due to Earth's rotation $\beta_L \approx 10^{-6}$. We express the measured frequency variation as a Fourier series
\be\label{FS}
\frac{\delta \nu}{\nu}=\frac{1}{8} \sum_{l,m,n} \left(C_{lmn} \cos\omega_{lmn}T+S_{lmn}\sin\omega_{lnm}T\right),
\ee
where the factor of $1/8$ is to simplify the Fourier coefficients $C_{lmn}, S_{lmn}$ and
\be
\omega_{lmn}=l\omega_t+m\omega_\oplus+n\Omega_\oplus.
\ee
The Fourier coefficients are listed in Tables \ref{SingleCoeffs} and \ref{DualCoeffs}. For the purpose of these tables, we use the definitions
\begin{eqnarray}\label{cspart}
c_Q^{\rm T}&=&c_{XX}^{\rm T}+c_{YY}^{\rm T}-2c_{ZZ}^{\rm T}, \nonumber \\
c_{-}^{\rm T}&=&c_{XX}^{\rm T}-c_{YY}^{\rm T}, \quad
c_{J}^w=|\varepsilon_{JKL}|c_{KL}^{\rm T}, \nonumber \\
c_{TJ}^{\rm T}&=&c_{TJ}^{\rm T}+c_{JT}^{\rm T},\quad
c_{TT}^{\rm T} =c_{TT}^{\rm T},
\end{eqnarray}
similar to Eq. (\ref{cs}) but without the factor of particle mass. They are related to the physical, single-particle coefficients $\tilde c$ by, e.g., $c_Q^{\rm T}=\sum_{\rm w} \eta^{\rm w}\tilde c_Q^{\rm w}$, and so on. In this equation, $\eta^{\rm w}=n^{\rm w}/m^{\rm T}$ is the number of particles of species w per mass of the composite object. For naturally abundant quartz, the numbers of electrons, protons and neutrons are quite similar and $\eta^{\rm e}\simeq \eta^{\rm p}\simeq \eta^{\rm n}\simeq 0.53/$GeV.

\begin{table*}
\centering
\caption{\label{SingleCoeffs} Signal components for one rotating crystal oscillator compared against a stationary reference that is not affected by the $c-$coefficients. Signal components suppressed by $\beta_\oplus^2$ and higher powers have been omitted. These coefficients are to be inserted in Eq. (\ref{FS}) and are multiplied by $1/8$ to give the frequency change.}
\begin{tabular}{ccc}\hline \hline
$l,m,n$ & cos & sin  \\ \hline
DC & $-2c^{\rm T}_Q(\sin^2\chi-2)$ & \\
0,0,1 & $2[-2c^{\rm T}_{TZ}\sin\eta\sin^2\chi+\cos\eta c^{\rm T}_{TY}(\sin^2\chi-2)]\beta_\oplus$ & $2(1+\cos^2\chi)c^{\rm T}_{TX}\beta_\oplus$ \\
0,1,-1 & $2\cos\chi\sin\eta\beta_\oplus c^{\rm T}_{TX}$ & $2\cos\chi(c^{\rm T}_{TZ}(1+\cos\eta)+c^{\rm T}_{TY}\sin\eta)\sin\chi\beta_\oplus$ \\
0,1,0 & $-2\sin\chi(2\cos\chi c^{\rm T}_Y+c^{\rm T}_{-}\sin\chi)$ & $-4\cos\chi\sin\chi c^{\rm T}_X$ \\
0,1,1 & $2\cos\chi c^{\rm T}_{TX}\sin\eta\sin\chi$ & $2\cos\chi[(\cos\eta-1)c^{\rm T}_{TZ}+c^{\rm T}_{TY}\sin\eta]\sin\chi\beta_\oplus$ \\
0,2,-1 & $(1+\cos\eta)(\cos^2\chi-1)\beta_\oplus c^{\rm T}_{TY}$ & $-(1+\cos\eta)(\cos^2\chi-1)c^{\rm T}_{TX}\beta_\oplus$ \\
0,2,0 & & $2(\cos^2\chi-1)c^{\rm T}_Z$ \\
0,2,1 & $(\cos\eta-1)(\cos^2\chi-1)c^{\rm T}_{TY}\beta_\oplus$ & $-(\cos\eta-1)(\cos^2\chi-1)c^{\rm T}_{TX}\beta_\oplus$ \\
2,-2,-1 & $(\cos\eta-1)(1-\cos\chi)^2c^{\rm T}_{TY}\beta_\oplus/2$ & $(\cos\eta-1)(\cos\chi-1)^2c^{\rm T}_{TX}\beta_\oplus/2$ \\
2,-2,0 & $c^{\rm T}_{-}(2-2\cos\chi-\sin^2\chi)+2(1-\cos\chi)c^{\rm T}_Y\sin\chi$ & $-(\cos\chi-1)^2c^{\rm T}_Z$  \\
2,-2,1 & $(1+\cos\eta)(1-\cos\chi)^2c^{\rm T}_{TY}\beta_\oplus/2$ & $(1+\cos\eta)(\cos\chi-1)^2c^{\rm T}_{TX}\beta_\oplus/2$ \\
2,-1,-1 & $(\cos\chi-1)\sin\eta\sin\chi c^{\rm T}_{TX}\beta_\oplus$ & $-(\cos\chi-1)[(\cos\eta-1)c^{\rm T}_{TZ}+\sin\eta c^{\rm T}_{TY}]\sin\chi\beta_\oplus$ \\
2,-1,0 & & $2(\cos\chi-1)c^{\rm T}_X$ \\
2,-1,1 & $(\cos\chi-1)c^{\rm T}_{TX}\sin\eta\beta_\oplus$ & $-(\cos\chi-1)[c^{\rm T}_{TZ}(1+\cos\eta)+c^{\rm T}_{TY}\sin\eta]\sin\chi\beta_\oplus$ \\
2,0,-1 & $(\cos\eta c^{\rm T}_{TY}-2c^{\rm T}_{TZ}\sin\eta)\sin^2\chi\beta_\oplus$ & $-(-1+\cos^2\chi)c^{\rm T}_{TX}\beta_\oplus$ \\
2,0,0 & $-2c^{\rm T}_Q\sin^2\chi$ & \\
2,0,1 & $-[\cos\eta(\cos^2\chi-1)c^{\rm T}_{TY}+2c^{\rm T}_{TZ}\sin\eta\sin^2\chi]\beta_\oplus$ & $(\cos^2\chi-1)c^{\rm T}_{TX}\beta_\oplus$\\
2,1,-1 & $(1+\cos\chi)\sin\eta\sin\chi\beta_\oplus c^{\rm T}_{TX}$ & $(1+\cos\chi)(c^{\rm T}_{TZ}(1+\cos\eta)+c^{\rm T}_{TY}\sin\eta)\sin\chi\beta_\oplus$ \\
2,1,0 & $-2(1+\cos\chi)\sin\chi c^{\rm T}_Y$ & $-2(1+\cos\chi)c^{\rm T}_X$ \\
2,1,1 & $(1+\cos\chi)c^{\rm T}_{TX}\sin\eta\beta_\oplus$ & $(1+\cos\chi)[(\cos\eta-1)c^{\rm T}_{TZ}+c^{\rm T}_{TY}\sin\eta]\sin\chi\beta_\oplus$ \\
2,2,-1 & $(1+\cos\eta)(1+\cos\chi)^2\beta_\oplus c^{\rm T}_{TY}/2$ & $-(1+\cos\eta)(1+\cos\chi)^2c^{\rm T}_{TX}\beta_\oplus/2$ \\
2,2,0 & $c^{\rm T}_{-}(2+2\cos\chi-\sin^2\chi)$ & $(1+\cos\chi)^2c^{\rm T}_Z$\\
2,2,1 & $(\cos\eta-1)(1+\cos\chi)^2c^{\rm T}_{TY}\beta_\oplus/2$ & $-(\cos\eta-1)(1+\cos\chi)^2c^{\rm T}_{TX}\beta_\oplus/2$ \\
\hline\hline
\end{tabular}
\end{table*}

\begin{table*}
\centering
\caption{\label{DualCoeffs} Signal components for an experiment with two rotating crystal oscillators compared against one another. Components suppressed by $\beta_\oplus^2$ and higher powers have been omitted. These coefficients are to be inserted in Eq. (\ref{FS}) and are multiplied by $1/8$ to give the frequency change.}
\begin{tabular}{ccc}\hline \hline
$l,m,n$ & cos & sin  \\ \hline
DC & $-4c^{\rm T}_Q\cos(2\theta)$ & \\
2,-2,-1 & $(\cos\eta-1)(\cos\chi-1)^2c^{\rm T}_{TY}\bo$ & $(\cos\eta-1)(\cos\chi-1)^2\cTX\bo$ \\
2,-2,0 & $2(\cos\chi-1)^2c^{\rm T}_M$ & 0 \\
2,-2,1 & $ (1 +\cos\eta) (-1 + \cos\chi)^2 \cTY \bo$ & $(1 + \cos\eta) (\cos\chi-1)^2 \cTX \bo$ \\
2,-1,-1 & $2 (\cos\chi-1) \cTX \sin\eta \sin\chi \bo$ & $2 (1 - \cos\chi) ((\cos\eta-1) \cTZ +
   \cTY \sin\eta) \sin\chi \bo$ \\
2,-1,0 & $4 (\cos\chi-1) \cY \sin\chi$ & $4 (\cos\chi-1) \cX \sin\chi$ \\
2,-1,1 & $2 (\cos\chi-1) \cTX \sin\eta \sin\chi \bo$ & $2 (1 - \cos\chi) (\cTZ + \cos\eta \cTZ +
   \cTY \sin\eta) \sin\chi \bo$ \\
2,0,-1 &  $2\sin^2\chi (\cos\eta \cTY - 2 \cTZ \sin\eta) \bo$ & $2\sin^2\chi \cTX \bo$ \\
2,0,0 & $-4\sin^2\chi c^{\rm T}_Q$ & 0 \\
2,0,1 & $2\sin^2\chi (\cos\eta \cTY - 2 \cTZ \sin\eta) \bo$ & $-2\sin^2\chi \cTX \bo$ \\
2,1,-1 & $2 (1 + \cos\chi) \cTX \sin\eta \sin\chi \bo$ & $2 (1 + \cos\chi) (\cTZ + \cos\eta \cTZ +
   \cTY \sin\eta) \sin\chi \bo$ \\
2,1,0 & $-4 (1 + \cos\chi) \cY \sin\chi$ & $-4 (1 + \cos\chi) \cX \sin\chi$ \\
2,1,1 &  $2 (1 + \cos\chi) \cTX \sin\eta \sin\chi \bo$ & $2 (1 + \cos\chi) ((-1 + \cos\eta) \cTZ +
   \cTY \sin\eta) \sin\chi \bo$ \\
2,2,-1 & $  (1 + \cos\eta) (1 + \cos\chi)^2 \cTY \bo$ & $- (1 +
   \cos\eta) (1 + \cos\chi)^2 \cTX \bo$ \\
2,2,0 & $2 c^{\rm T}_- (1 + \cos\chi)^2$ & $2 (1 + \cos\chi)^2 \cZ$ \\
2,2,1 & $ (\cos\eta-1) (1 + \cos\chi)^2 \cTY \bo$ & $ (1 -
   \cos\eta) (1 + \cos\chi)^2 \cTX \bo$ \\
\hline\hline
\end{tabular}
\end{table*}

\section{Preliminary experiment with room-temperature oscillators}

Optimized phonon-sector experiments will be able to improve bounds in all fermion sectors of the SME, as we will see below. For our current experiment, however, we will focus on the neutron sector. Existing limits on the proton and electron-sector coefficients are at levels somewhat below the sensitivity of this preliminary experiment \cite{datatables}.

\subsection{Previous neutron-sector limits}
The most sensitive experiments to determine limits on the $\tilde c^{\rm n}$ for neutrons are based on magnetometry \cite{Prestage,Lamoreaux,Chupp,KosteleckyLane,Brown,Smiciklas}. In particular, a Neon/Rubidium/Potassium co-magnetometer has been used, which simultaneously sense the influence of background magnetic fields and the signal for Lorentz violation. This bounds the four spatial components $\tilde c^n_{J}, \tilde c$ to the very low level of $10^{-29}$ \cite{Smiciklas}. Limits on the fifth, $c^n_Q$, are not available from this experiment, but are available from tests of the weak equivalence principle \cite{KosteleckyTasson} and astrophysics \cite{Altschul}.
%Ad Hohensee: define clean in a better way (ref3)
% Added leading definition of 'clean', and removed ``clean''.  Switched 'available' to 'possible' -Hohensee
Without making untested assumptions about the character or degree to which Lorentz symmetry is broken in other sectors of the SME, the best laboratory limit on $c^{n}_{Q}$ is $|\tilde c^n_Q|<10^{-8}$\,GeV \cite{KosteleckyTasson}. Assuming that the $\alpha a_{\rm eff}$-coefficients vanish, an improved limit of $10^{-11}$\,GeV is possible \cite{KosteleckyTasson}. Astrophysics studies of the stability of cosmic ray protons yields $|\tilde c_Q|<2\times 10^{-13}$\,GeV, $|\tilde c_{TJ}|< 5\times 10^{-14}$\,GeV \cite{Altschul}. The temporal $\tilde c^n_{TT}$ have been measured in atom interferometry \cite{redshiftPRL}.

%pulsar timing [77]

\subsection{Setup}
\begin{figure}
\centering
\epsfig{file=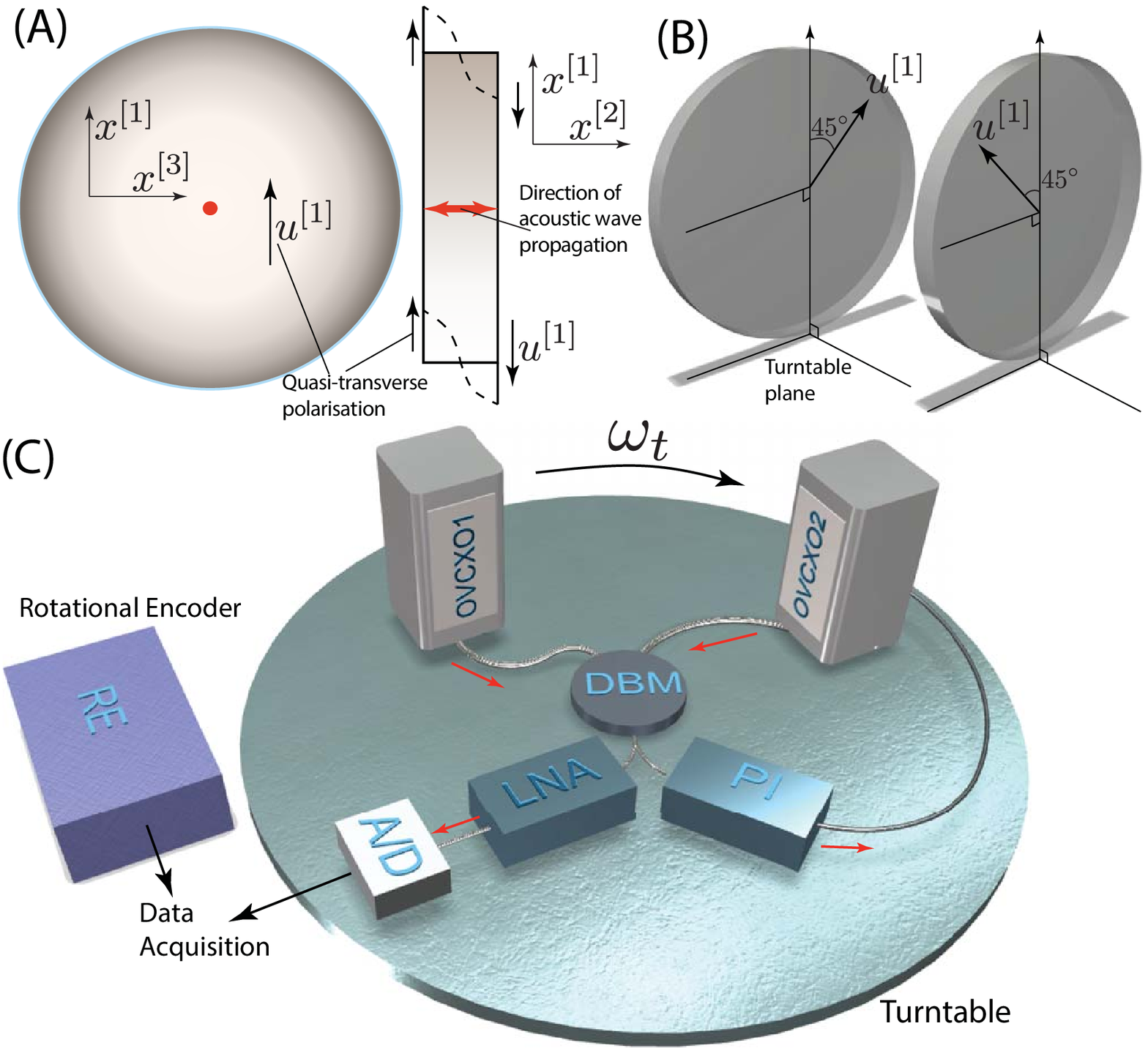,width=0.5\textwidth}
\caption{\label{TwoOsc} Schematic of the room temperature experiment. (A) Crystal blank showing coordinates $x^{[1-3]}$, the direction of propagation and the direction of the displacement of the shear mode. (B) Orientation of a shear relative to the turntable plane. (C) Frequency comparison on the turntable: The phase between two 10-MHz ovenized, voltage-controlled crystal oscillators (OVCXO) is detected by homodyne detection using a double-balanced mixer (DBM). Phase-lock with a PI feedback controller keeps the DBM operating near zero output voltage. The feedback is very slow, leaving the oscillators essentially free-running on the timescale of the turntable rotation rate and its harmonics. The DBM output is amplified by a low noise amplifier (LNA) and acquired after Analog-to-Digital (A/D) conversion. All components are enclosed in a cylindrical dual-layer $\mu$-metal magnetic shield (not shown).}
\end{figure}

Our experiment (Fig. \ref{TwoOsc}) uses active rotation at a frequency of $\om_t=2\pi\times 0.36$\,Hz on a precision air-bearing turntable. Relative to experiments based solely on Earth's rotation, this increase the signal frequencies and thus allows us to suppress the drift of the oscillators due, e.g., to ageing or temperature instability. The turntable (Professional Instruments, model 10R-606) is specified to $0.1\mu$rad tilt of the rotation axis and $<25\,$nm radial and axial wobble, and has a specified stiffness of 10\,Nm/$\mu$radian. We use two oscillators that are rotating on the turntable and that are directly compared on the turntable. This avoids the need to bring the signals in or out of the turntable. (The target accuracy of $10^{-13}$ out of 10\,MHz requires us to detect phase modulations of microradian-size; any modulations introduced when transmitting the signal from the turntable to the stationary laboratory frame would be synchronized with the putative signal, and none of the available methods can be trusted to not introduce tiny phase or amplitude modulations).  The oscillators (Stanford Research Systems SC-10) are signal generators classified as ovenized voltage-controlled crystal oscillators (OCXO) based on quartz SC-cut BAW resonators. The oscillators are specified to an Alan variance of $2\times 10^{-12}$ at 1 second averaging time. All components are highly reliable and covered with a mu-metal shield, allowing the experiment to take uninterrupted data over long stretches of time.

Directly comparing the frequency of the oscillators via an available frequency counter is limited to a resolution of about $10^{-11}$ in one second by the $\sim 100-$ps timing resolution of the device. Much higher resolution can be achieved by using a double-balanced mixer (Mini-Circuits RPD-1) to measure the phase difference between the oscillators. We use the mixer's internal signal transformers to provide galvanic isolation between the quartz oscillators and the direct-current (dc) circuits (the RPD-1 allows the three ports to have separate grounds) to avoid dc signal errors through ground loops, given the large supply current of the quartz ovens. The output signal of the mixer is pre-amplified 1000 times and the resulting voltage $U$ is digitized on the turntable. The digital signal is brought out of the turntable via a universal serial bus (USB) connection through slip-ring contacts. Power at 15\,V is also supplied via sliprings.

\begin{figure}
\centering
\epsfig{file=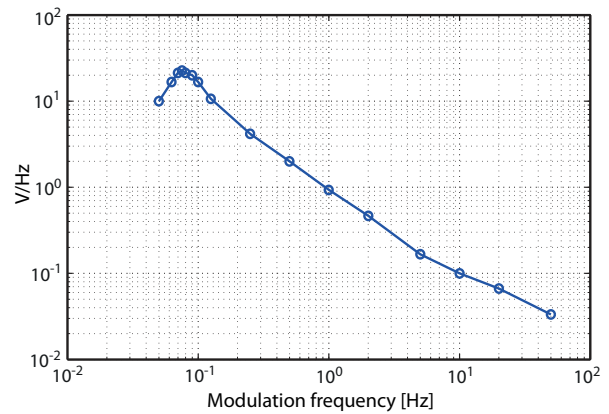,width=0.45\textwidth}
\caption{\label{ConversionEfficiency} Conversion efficiency of the mixer as frequency discriminator, measured with the phase-lock loop closed by inserting a signal having a known frequency modulation.}
\end{figure}

On timescales much longer than the rotation period of our turntable, we phase-lock the oscillators together so that the mixer may always operate close to 90$^\circ$ phase difference, i.e., near-zero output signal. The effective frequency-to-voltage conversion factor measured at the mixer output is thus zero at extremely low frequencies, where any voltages are removed by the feedback loop; at high frequencies, where the feedback is ineffective, the factor is given by purely by the mixer itself. We measure the conversion efficiency of the mixer as a frequency discriminator by replacing one of the quartz oscillators with a digital synthesizer which provides a known frequency modulation. Fig. \ref{ConversionEfficiency} shows the measured response function. At our signal frequency of $2\om_t\sim 2\pi \times 0.76\,$Hz, we obtain $\delta \nu=\Delta U/(1.3\,$V/Hz).

The turntable is driven by an unregulated dc motor. Even small changes of the rotation rate accumulate to a large angle offset over time. We therefore use a lightgate as a rotation encoder that delivers one pulse per turn to the computer, re-setting the angle scale of the turntable rotation. The computer then interpolates linearly assuming a constant rotation rate during one turn.

\subsection{Results}
\begin{figure}
\centering
\epsfig{file=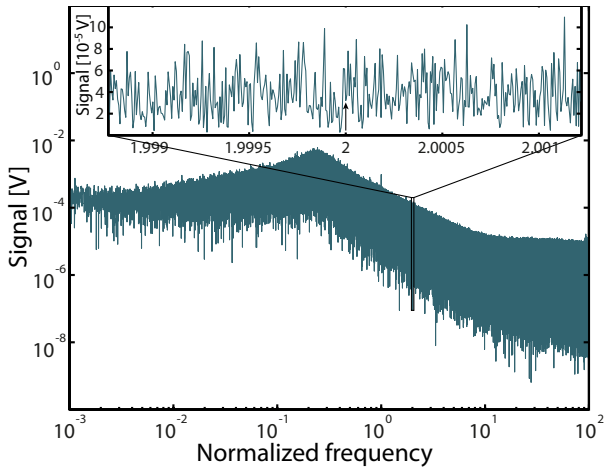,width=0.45\textwidth}
\caption{\label{FT} Fourier transform of 120 hours of data. Frequency is measured in multiples of the turntable rotation frequency.}
\end{figure}

The system proved to be extremely reliable and capable of unattended operation. Fig. \ref{FT} shows the amplitude Fourier transform of 120.0 hours of data (about 164,000 turntable rotations). Zooming into the region close to the expected signals around $2\om_t$ reveals sine and cosine amplitudes that are normally distributed with a standard deviation of $\sigma^2=\langle A_c^2\rangle=32\,\mu$V after amplification. The measured signal at $2\om_t$ is $-26\,\mu$V. This corresponds to $(-26 \pm 32)$\,nV at the mixer output and thus $\delta \nu/\nu= (-2.0 \pm 2.4)\times 10^{-15}$, see Fig. \ref{ConversionEfficiency}.

The signal for Lorentz violation (Tab. \ref{DualCoeffs}) has components at various frequencies around $2\om_t$. At our present accuracy, we restrict our analysis to the effect of $c_Q^{\rm Q}$, which causes a signal proportional to $\cos(2\om_t T)$. For its amplitude, we find $\delta \nu/\nu=\frac12 \sin^2\chi c^{\rm Q}_Q\simeq 0.31 c^{\rm Q}_Q$. In the experiment, however, the axes of the oscillators were oriented 45$^\circ$ relative to the rotation axis, which we take into account by a factor of $\sin 45^\circ$. We thus find $c^{\rm Q}_Q=(-0.9\pm 1.1)\times 10^{-14}$ on the effective coefficient for naturally abundant quartz, which translates into a limit of $\tilde c^{\rm n}_Q=(-1.8 \pm 2.2)\times 10^{-14}$\,GeV on the neutron-sector coefficient.

\begin{figure}
\centering
\epsfig{file=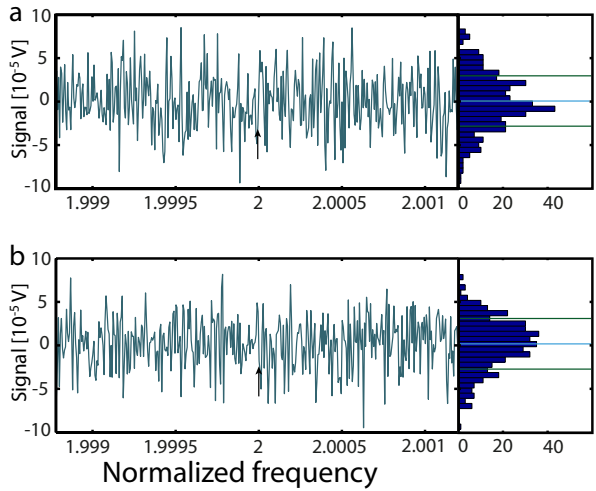,width=0.45\textwidth}
\caption{\label{SinCos} Cosine (a) and sine (b) Fourier transform around $2\om_t$. The arrow points out the putative signal at $2\om_t$.}
\label{Timetrace}
\end{figure}

Systematic effects of quartz oscillators such as aging, temperature fluctuations and thermal hysteresis, acceleration, magnetic fields, power supply voltage, load impedance, electric fields, ionizing radiation, and ground loops, are well-understood. At our current resolution, most systematics are negligible so here we discuss the largest two effects: We measured the acceleration sensitivity of our quartz oscillators by inverting them relative to Earth's gravitational acceleration $g$. For the most sensitive axis, we find $\delta \nu\sim 20$\,mHz$/2g$. The turntable wobble is specified to be less than 25\,nm radially and axially. If we conservatively assume that this wobble contributes a $2\om_t$ frequency component (in reality, the energy of the wobble is likely spread out over many Fourier components), the corresponding acceleration is 25\,nm$\times 4\om_t^2\sim 0.14\times 10^{-6}g$, which produces frequency changes of $2.8\,$nHz. Changing magnetic fields induce voltages into our wiring. Assuming 1\,Gauss and an enclosed area of 1\,cm$^2$ at the turntable frequency $2\om_t$ (conservatively assuming that all the magnetic field will contribute to the second harmonic of the turn table rate), we obtain an induced voltage of $\sim 40$\,nV, comparable to our signal size. For this reason, we enclose the entire setup up to and including the amplifier in a two-layer mu-metal shield, which should reduce this influence at least $\sim 100-$ fold.

%\begin{tabular}{cccc}\hline\hline
%Effect	& input      	& assumption     	& $\delta \nu/\nu$ \\ \hline
%Radiation & $10^{-11}$/rad \cite{Oscilent} & $10^{-5}$ rad/year & $10^{-24}$ \\
%Acceleration & $10^{-12}/g$  & $1.4\times 10^{-7}g$ & $2.8\times 10^{-16}$ \\ %\hline\hline
%\end{tabular}

\subsection{Cryogenic experiment}

\begin{figure}[t!]
    \centering
\epsfig{file=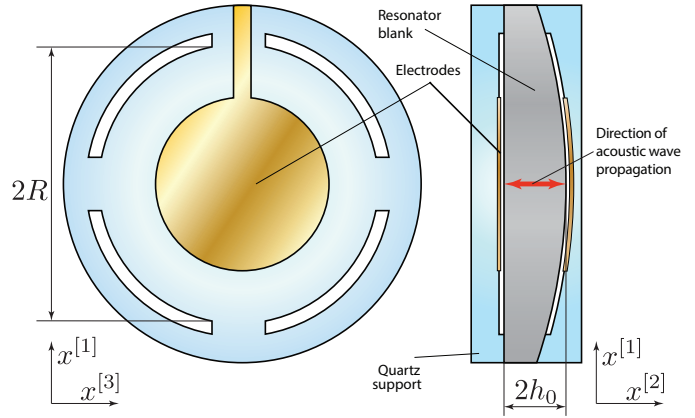,width=0.5\textwidth}
    \caption{Schematic view of a BVA (Bo\^{i}tier \`{a} Vieillissement Am\'{e}lior\'e, Enclosure with Improved Aging) BAW quartz resonator. The vibrating quartz body, resonator blank, is held by a quartz support deposited electrodes. The resonator is fabricated in a plano-convex geometry that traps vibration in the disk center. Electrodes are deposited on the support and separated from the plank by a small vacuum gap.}
    \label{gapp}
\end{figure}

The quartz bulk acoustic wave (BAW) technology provides the most stable oscillators in the medium and high frequency range ($1-50$~MHz) between 1 and 30 seconds of averaging time. Such oscillators are also the most stable macroscopic mechanical harmonic oscillators, with fractional frequency stabilities as low as $2.5\times10^{-14}$ \cite{Salzenstein:2010aa} for room temperature devices. Over the last decade there has been no major improvement in quartz oscillator performance at room temperature, mainly due to the quartz resonator self-noise. For this reason, the electrodeless (or BVA)\cite{1537081} BAW quartz resonators (see Fig.~\ref{gapp}) have been investigated for cryogenic operation. These investigations reveal extremely high values of the quality factors exceeding $10^9$ \cite{galliou:091911,ours} as well as an ability to operate at high overtones \cite{quartzPRL} providing a new platform for many physical experiments \cite{ScRep,Goryachev:2014aa}, for example detection of high frequency gravitational waves\cite{GW} and cooling a macroscopic object to its ground state for tests of fundamental physics \cite{quartzPRL,Pikovski}. Table~\ref{modesT} gives values of quality factor for some overtones measured in a 4K environment.

\begin{table}[t]
\caption{Values of $Q$-factors for some overtones of cryogenic ($4$K) BAW resonators.}
\centering
\begin{tabularx}{\columnwidth}{XXX|XXX}
\hline
\hline
X$_{n,m,p}$ & $f_r,$ MHz & $Q,10^{8}$ & X$_{n,m,p}$ & $f_r,$ MHz & $Q,10^{7}$ \\
\hline
C$_{3,0,0}$ & 4.99 & 0.4 & A$_{23,0,0}$ & 72.22 & 5.01\\
C$_{5,0,0}$ & 8.39 & 1.1 & A$_{25,0,0}$ & 78.51 & 2.98\\
B$_{3,0,0}$ & 5.51 & 0.5 & A$_{27,0,0}$ & 84.78 & 41.2\\
B$_{5,0,0}$ & 8.39 & 0.6 & A$_{33,0,0}$ & 103.6 & 42.3\\
B$_{5,m,p}$ & 9.15 & 1.2 & A$_{37,0,0}$ & 116.2 & 49.6\\
B$_{5,m,p}$ & 9.25 & 2.6 & A$_{43,0,0}$ & 135.0 & 35.6\\
A$_{3,0,0}$ & 9.37 & 0.6 & A$_{45,0,0}$ & 141.2 & 33.5\\
A$_{5,0,0}$ & 15.97 & 3 & A$_{47,0,0}$ & 147.5 & 20.0\\
A$_{15,0,0}$ & 47.11 & 19.3 & A$_{55,0,0}$ & 172.6 & 49.6\\
\hline
\end{tabularx}
\label{modesT}
\end{table}

Such a significant increase of the quality factor may result in reduction of the oscillator fractional frequency stability. Assuming that the  dominant flicker noise of the resonator at 1\,Hz from the carrier ($S_\phi(1\text{Hz})\sim-130\frac{\text{dBrd}^2}{\text{Hz}}$) does not change between cryogenic and room temperatures, the Allan deviation of a cryogenic source may be estimated to achieve a level of
\begin{equation}
    \label{B003SF}
\displaystyle  \sigma_y = \frac{1}{2Q}\sqrt{2\ln{2}S_\phi(1\text{Hz})}\sim 2\times 10^{-16}.
\end{equation}

BAW resonator ageing is a systematic drift of its resonant frequency that can be typically observed at long averaging times. This process usually gives a slope of $\tau^{1}$ (where $\tau$ is the integration time) in the Alan Deviation curve dominating over $\tau^{1/2}$ law resulting from thermal fluctuations for averaging times over $10^3$ seconds. The ageing process can be caused by a number of effects primarily related to manufacturing. This process is the most prominent during the first months of the oscillator continuous operation, which gradually decreases during that time. For the current experiment, the ageing is about $2\times10^{-10}$pp/day. For ultra-stable oscillators this can be reduced to $3\times10^{-12}$~pp/day or $1\times10^{-9}$~pp/year after at least 90 days of continuous operation.

Another source of stability improvement is associated with the relation between the flicker and white noise in typical BAW oscillators. While the white noise is connected to the signal-to-noise ratio and could be thus reduced by increasing the oscillation power, the flicker noise drops with decreasing power. This situation results in a compromise between the mid term stability (flicker noise region) and short term stability (white noise region). At cryogenic temperatures the white noise is naturally reduced according to the Nyquist relation, thus giving more room for flicker noise improvement by oscillator power reduction. The Nyquist noise limit for BAW resonators has been recently demonstrated at liquid helium temperatures\cite{Goryachev:2014ab} and unequivocally demonstrates the drop in this limit.

Nevertheless, practical realization of such a cryogenic BAW-clock is associated with technical difficulties \cite{Goryachev:2012jx,Goryachev:2013ly}. So far only moderate long temperature stability improvement has been demonstrated \cite{SunFr,Goryachev:2013ly}. The main problem is the absence of a frequency-temperature turn-over point giving rise to significant fluctuations. Additionally, the absence of reliable low temperature components at the medium and high frequency range makes the oscillator design a challenging problem. Whereas, the first problem may be overcome by a design of a special cut for cryogenic temperatures, the second is solvable by shifting from semiconductor to superconductor technology. Furthermore, for realizing Lorentz violation experiments, which utilize two oscillators, one just needs to match temperature coefficients of two orthogonally orientated resonators, so as to read out a stable beat frequency, in a similar way to the cryogenic sapphire oscillator tests in the photon sector\cite{Stanwix}. This may relax the requirements for a turnover point for these types of measurements.

\subsection{Potential sensitivity}
Given our expected stability of the cryogenic source given in Eq. (12), we can expect a four orders of magnitude improvement in sensitivity, compare to the room temperature measurement, thus a cryogenic quartz oscillator experiment should be able to test Lorentz invariance with a sensitivity to fractional frequency changes of $10^{-19}$. We now estimate the limits that can be derived from such a phonon-sector experiment to the coefficients of the Fermions and photons in the SME. Optimistically, perhaps the performance can be improved by a further order of magnitude or more.

The estimates are based on the non-relativistic single-particle hamiltonian describing Lorentz violation in the SME, see, e.g., Eq. (4) in \cite{KosteleckyLane}. This hamiltonian contains constant terms, terms proportional to the spin $\sigma^j$, and terms proportional to momentum $p^j$. These terms are unmeasurable as they are either constant, or average out over a cycle of the acoustic oscillation. The term proportional to $p^j\sigma^k$ averages out as well, but can perhaps be measured by applying an oscillating magnetic field at the same frequency as the acoustic oscillation, which would periodically polarize the spin and thus cause a nonzero $\langle p^j\sigma^k\rangle$. We will not consider this. Measurable signals arise from two terms in the hamiltonian. The one proportional to $p^j p^k$ allows measuring the $\tilde c$ coefficients, as discussed in detail above. Using spin-polarized materials, the term proportional to $p^jp^k\sigma^l$ can be measurable.

Table \ref{potlim} shows the order of magnitude of sensitivities from a phonon-sector experiment with a frequency sensitivity of $10^{-18}$. For each entry, all other coefficients for Lorentz violation are assumed to vanish. An asterisk highlights coefficients for which the phonon-sector experiment can provide improved bounds, based on comparison with the maximum sensitivities table in the 2015 edition of the Data Tables on Lorentz and CPT violation \cite{datatables}. Two asterisks highlight entries where the phonon-sector experiment would provide the first bound on a parameter, based on the same comparison.

The $\tilde b_J$ coefficients enter from the $p^j p^k \sigma^l$ term. For the estimate, we assume that $\langle \sigma \rangle \sim $10\% of all electron spins and $10^{-6}$ of all nuclear spins have been polarized. The $\tilde b_T$ coefficients enters through the same mechanism, suppressed by a factor of $\beta_\oplus$, the Earth's orbital velocity. The $\tilde d$ coefficients enter through spin polarization, similar to the $\tilde b$'s.

The fact that bindings in crystals are electromagnetic results in an influence of photon-sector coefficients \cite{ResSME,H2SME,TLVMC}. This is expected to result in leading-order signals in phonon-sector experiments as reflected in the table. A detailed calculation would have to be specific for the material used in the experiment. This is beyond the scope of this paper.

\begin{table}
\centering
\caption{\label{potlim} Potential order-of-magnitude sensitivities from phonon-sector experiments, assuming a $\delta \nu/\nu=10^{-19}$ resolution. Tabulated is $\log_{10}$ of the expected bound in GeV. Limits in parentheses assume a spin-polarized solid. For entries marked with a *, the sensitivity of phonon-sector experiments are equal or better than the maximal sensitivity listed in the 2015 edition of \cite{datatables} for at least one component. For entries highlighted by two asterisks (**), no limit is available in \cite{datatables} on at least one component.}
\begin{tabular}{ccccc}\hline\hline
Coefficient & Electron & Proton & Neutron & Photon \\ \hline
$\tilde b_J$ & (-18) & (-13) & (-13) & \\
$\tilde b_T$ & (-14) & (-9)* & (-9) & \\
$\tilde c_Q$ & -18* & -18 & -18* &  \\
$\tilde c_{-}, \tilde c_{J}$ & -18 & -18 & -18 &  \\
$\tilde c_{TJ}$ & -14 & -14 & -14* & \\
$\tilde c_{TT}$ & -10 & -10 & -10 & \\
$\tilde d_+,\tilde d_Q$ & (-18) & (-13)* & (-13) & \\
$\tilde d_-, \tilde d_{JK}$ & (-18) & (-13)** & (-13) & \\
$\tilde d_J$ & (-18) & (-13)** & (-13)** &    \\
$\tilde H_{JT}$ & & & & \\
$\tilde g_c, \tilde g_Q, \tilde g_- , \tilde g_T$ & & & & \\
$\tilde g_{TJ}, \tilde g_{JK}, \tilde g_{DJ}$ & & & & \\
$\tilde \kappa_{e-}$ & & & & -18* \\
$\tilde \kappa_{o+}$ & & & & -14* \\
$\tilde\kappa_{e+}$ & & & & \\
$\tilde\kappa_{o-}$ & & & & \\
$\tilde \kappa_{tr}$ & & & & -10 \\
\hline\hline
\end{tabular}
\end{table}

Like the photon terms, the coefficients entering the equations of motion of the valence electrons will modify the bindings, which will perhaps provide additional signals for the electron terms of the SME. These signals will be roughly proportional to the {\em relative} change in inertial mass and thus given by the $c^{\rm e}$ coefficients. Their effect may potentially be quite strong. An electron coefficient of, e.g., $\tilde c^{\rm e}_Q=10^{-20}\,$GeV corresponds to $c^{\rm e}_{XX}+c^{\rm e}_{YY}-2c^{\rm e}_{ZZ}\simeq 2\times 10^{-17}$ because $1/m^{\rm e}\sim 2000\,$GeV$^{-1}$, and might thus be measurable. As above, the detailed analysis of specific crystals is beyond the scope of the paper.

\section{Summary and outlook}

We have presented a new method for testing Lorentz symmetry, frequency comparisons between quartz crystal oscillators. %Such oscillators are commercially available and highly reliable.
While their stability today is surpassed by atomic clocks (especially optical clocks), many tests of Lorentz symmetry are not limited by signal-to-noise, but often by systematic effects from wobble and tilt of the turntable, and the ability to take data over long stretches of time. Quartz oscillators are compact, simple to apply and to shield from environmental influences. Their low acceleration sensitivity makes them relatively immune to wobble and tilt. Maintenance-free operation allows for long-term data taking which helps to make up for the reduced stability. As a demonstration, we have improved the laboratory limit on the neutron-$c_Q$ coefficient by six orders of magnitude, surpassing even current astrophysics bounds. Currently cryogenic oscillators are under development at University of Western Australia and FEMTO-ST, and promise strong improvements in stability and the sensitivity to Lorentz violating coefficients.

By analogy to photon sector experiments, we believe our method can be strongly improved.  Photon sector experiments have gained four orders of magnitude in sensitivity over the last twelve years, through higher quality factors resonance and cryogenic operation. A cryogenic version of our experiment may increase the quality factor of the resonance about 10,000 fold and may strongly reduce the temperature coefficient of the oscillators. We thus estimate that three to four orders of magnitude improvement to $\sim 10^{-18}$ frequency resolution are realistic. This new technology may also lead to milligram-scale mechanical oscillators at the quantum limit and may see a new brand of ultra-stable oscillators. At this sensitivity, the experiment will be able to improve the bounds on several Fermion- and photon sector coefficients, including several coefficients that are not bounded today. Likewise, a detailed theoretical analysis of specific materials might reveal large additional sensitivities to the electron $c$- coefficients.

%Taking data over a year would allow us to independently measure all coefficients for Lorentz violation in Tab. \ref{DualCoeffs} and \ref{SingleCoeffs}, as in . %The experiment is in principle sensitive to any Lorentz violation that changes the inertial masses of protons and neutrons. %With a further 10-fold improvement in sensitivity, it will be able to set the best limits on the $c_{TJ}^n$ coefficients for neutrons.
%The influence of anisotropic inertial masses of electrons is suppressed by their lower mass.
%Other applications of crystal oscillators might include the search for topological dark matter \cite{PospelovDerevianko}.

\acknowledgements We thank Justin Brown and Alan Kosteleck\'y for discussions. This work was supported by the David and Lucile Packard foundation, the Australian Research Council Grant No. CE110001013 and DP130100205, the Austrian Science Fund (FWF): J3680, and was performed under the auspices of the U.S. Department of Energy by Lawrence Livermore National Laboratory under Contract DE-AC52-07NA27344.

\bibliography{biblio}

\begin{thebibliography}{83}
\expandafter\ifx\csname natexlab\endcsname\relax\def\natexlab#1{#1}\fi
\expandafter\ifx\csname bibnamefont\endcsname\relax
  \def\bibnamefont#1{#1}\fi
\expandafter\ifx\csname bibfnamefont\endcsname\relax
  \def\bibfnamefont#1{#1}\fi
\expandafter\ifx\csname citenamefont\endcsname\relax
  \def\citenamefont#1{#1}\fi
\expandafter\ifx\csname url\endcsname\relax
  \def\url#1{\texttt{#1}}\fi
\expandafter\ifx\csname urlprefix\endcsname\relax\def\urlprefix{URL }\fi
\providecommand{\bibinfo}[2]{#2}
\providecommand{\eprint}[2][]{\url{#2}}

\bibitem[{\citenamefont{Colladay and
  Kosteleck{\'y}}(1997)}]{ColladayKostelecky}
\bibinfo{author}{\bibfnamefont{D.}~\bibnamefont{Colladay}} \bibnamefont{and}
  \bibinfo{author}{\bibfnamefont{V.~A.} \bibnamefont{Kosteleck{\'y}}},
  \bibinfo{journal}{Physical Review D} \textbf{\bibinfo{volume}{55}},
  \bibinfo{pages}{6760} (\bibinfo{year}{1997}),
  \urlprefix\url{http://link.aps.org/doi/10.1103/PhysRevD.55.6760}.

\bibitem[{\citenamefont{Kosteleck{\'y} and Mewes}(2006)}]{KosteleckyMewes}
\bibinfo{author}{\bibfnamefont{V.~A.} \bibnamefont{Kosteleck{\'y}}}
  \bibnamefont{and} \bibinfo{author}{\bibfnamefont{M.}~\bibnamefont{Mewes}},
  \bibinfo{journal}{Physical Review Letters} \textbf{\bibinfo{volume}{97}},
  \bibinfo{pages}{140401} (\bibinfo{year}{2006}),
  \urlprefix\url{http://link.aps.org/doi/10.1103/PhysRevLett.97.140401}.

\bibitem[{\citenamefont{Kosteleck{\'y} and Mewes}(2002)}]{KosteleckyMewesPRD}
\bibinfo{author}{\bibfnamefont{V.~A.} \bibnamefont{Kosteleck{\'y}}}
  \bibnamefont{and} \bibinfo{author}{\bibfnamefont{M.}~\bibnamefont{Mewes}},
  \bibinfo{journal}{Physical Review D} \textbf{\bibinfo{volume}{66}},
  \bibinfo{pages}{056005} (\bibinfo{year}{2002}),
  \urlprefix\url{http://link.aps.org/doi/10.1103/PhysRevD.66.056005}.

\bibitem[{\citenamefont{Herrmann et~al.}(2009)\citenamefont{Herrmann, Senger,
  M{\"o}hle, Nagel, Kovalchuk, and Peters}}]{Achim}
\bibinfo{author}{\bibfnamefont{S.}~\bibnamefont{Herrmann}},
  \bibinfo{author}{\bibfnamefont{A.}~\bibnamefont{Senger}},
  \bibinfo{author}{\bibfnamefont{K.}~\bibnamefont{M{\"o}hle}},
  \bibinfo{author}{\bibfnamefont{M.}~\bibnamefont{Nagel}},
  \bibinfo{author}{\bibfnamefont{E.~V.} \bibnamefont{Kovalchuk}},
  \bibnamefont{and} \bibinfo{author}{\bibfnamefont{A.}~\bibnamefont{Peters}},
  \bibinfo{journal}{Physical Review D} \textbf{\bibinfo{volume}{80}},
  \bibinfo{pages}{105011} (\bibinfo{year}{2009}),
  \urlprefix\url{http://link.aps.org/doi/10.1103/PhysRevD.80.105011}.

\bibitem[{\citenamefont{Stanwix
  et~al.}(2005{\natexlab{a}})\citenamefont{Stanwix, Tobar, Wolf, Susli, Locke,
  Ivanov, Winterflood, and van Kann}}]{Stanwix}
\bibinfo{author}{\bibfnamefont{P.~L.} \bibnamefont{Stanwix}},
  \bibinfo{author}{\bibfnamefont{M.~E.} \bibnamefont{Tobar}},
  \bibinfo{author}{\bibfnamefont{P.}~\bibnamefont{Wolf}},
  \bibinfo{author}{\bibfnamefont{M.}~\bibnamefont{Susli}},
  \bibinfo{author}{\bibfnamefont{C.~R.} \bibnamefont{Locke}},
  \bibinfo{author}{\bibfnamefont{E.~N.} \bibnamefont{Ivanov}},
  \bibinfo{author}{\bibfnamefont{J.}~\bibnamefont{Winterflood}},
  \bibnamefont{and} \bibinfo{author}{\bibfnamefont{F.}~\bibnamefont{van Kann}},
  \bibinfo{journal}{Physical Review Letters} \textbf{\bibinfo{volume}{95}},
  \bibinfo{pages}{040404} (\bibinfo{year}{2005}{\natexlab{a}}),
  \urlprefix\url{http://link.aps.org/doi/10.1103/PhysRevLett.95.040404}.

\bibitem[{\citenamefont{Eisele et~al.}(2009)\citenamefont{Eisele, Nevsky, and
  Schiller}}]{Schiller}
\bibinfo{author}{\bibfnamefont{C.}~\bibnamefont{Eisele}},
  \bibinfo{author}{\bibfnamefont{A.~Y.} \bibnamefont{Nevsky}},
  \bibnamefont{and} \bibinfo{author}{\bibfnamefont{S.}~\bibnamefont{Schiller}},
  \bibinfo{journal}{Physical Review Letters} \textbf{\bibinfo{volume}{103}},
  \bibinfo{pages}{090401} (\bibinfo{year}{2009}),
  \urlprefix\url{http://link.aps.org/doi/10.1103/PhysRevLett.103.090401}.

\bibitem[{\citenamefont{M{\"u}ller}(2005)}]{TLVMC}
\bibinfo{author}{\bibfnamefont{H.}~\bibnamefont{M{\"u}ller}},
  \bibinfo{journal}{Physical Review D} \textbf{\bibinfo{volume}{71}},
  \bibinfo{pages}{045004} (\bibinfo{year}{2005}),
  \urlprefix\url{http://link.aps.org/doi/10.1103/PhysRevD.71.045004}.

\bibitem[{\citenamefont{Kosteleck\'y and Lane}(1999)}]{KosteleckyLane}
\bibinfo{author}{\bibfnamefont{V.~A.} \bibnamefont{Kosteleck\'y}}
  \bibnamefont{and} \bibinfo{author}{\bibfnamefont{C.~D.} \bibnamefont{Lane}},
  \bibinfo{journal}{Phys. Rev. D} \textbf{\bibinfo{volume}{60}},
  \bibinfo{pages}{116010} (\bibinfo{year}{1999}),
  \urlprefix\url{http://link.aps.org/doi/10.1103/PhysRevD.60.116010}.

\bibitem[{\citenamefont{Wolf et~al.}(2006)\citenamefont{Wolf, Chapelet, Bize,
  and Clairon}}]{WolfPRD2006}
\bibinfo{author}{\bibfnamefont{P.}~\bibnamefont{Wolf}},
  \bibinfo{author}{\bibfnamefont{F.}~\bibnamefont{Chapelet}},
  \bibinfo{author}{\bibfnamefont{S.}~\bibnamefont{Bize}}, \bibnamefont{and}
  \bibinfo{author}{\bibfnamefont{A.}~\bibnamefont{Clairon}},
  \bibinfo{journal}{Physical Review Letters} \textbf{\bibinfo{volume}{96}},
  \bibinfo{pages}{060801} (\bibinfo{year}{2006}),
  \urlprefix\url{http://link.aps.org/doi/10.1103/PhysRevLett.96.060801}.

\bibitem[{\citenamefont{Flambaum et~al.}(2009)\citenamefont{Flambaum, Lambert,
  and Pospelov}}]{Flambaum2009}
\bibinfo{author}{\bibfnamefont{V.}~\bibnamefont{Flambaum}},
  \bibinfo{author}{\bibfnamefont{S.}~\bibnamefont{Lambert}}, \bibnamefont{and}
  \bibinfo{author}{\bibfnamefont{M.}~\bibnamefont{Pospelov}},
  \bibinfo{journal}{Physical Review D} \textbf{\bibinfo{volume}{80}},
  \bibinfo{pages}{105021} (\bibinfo{year}{2009}),
  \urlprefix\url{http://link.aps.org/doi/10.1103/PhysRevD.80.105021}.

\bibitem[{\citenamefont{Can{\`e} et~al.}(2004)\citenamefont{Can{\`e}, Bear,
  Phillips, Rosen, Smallwood, Stoner, Walsworth, and Kosteleck{\'y}}}]{Crane}
\bibinfo{author}{\bibfnamefont{F.}~\bibnamefont{Can{\`e}}},
  \bibinfo{author}{\bibfnamefont{D.}~\bibnamefont{Bear}},
  \bibinfo{author}{\bibfnamefont{D.~F.} \bibnamefont{Phillips}},
  \bibinfo{author}{\bibfnamefont{M.~S.} \bibnamefont{Rosen}},
  \bibinfo{author}{\bibfnamefont{C.~L.} \bibnamefont{Smallwood}},
  \bibinfo{author}{\bibfnamefont{R.~E.} \bibnamefont{Stoner}},
  \bibinfo{author}{\bibfnamefont{R.~L.} \bibnamefont{Walsworth}},
  \bibnamefont{and} \bibinfo{author}{\bibfnamefont{V.~A.}
  \bibnamefont{Kosteleck{\'y}}}, \bibinfo{journal}{Physical Review Letters}
  \textbf{\bibinfo{volume}{93}}, \bibinfo{pages}{230801}
  (\bibinfo{year}{2004}),
  \urlprefix\url{http://link.aps.org/doi/10.1103/PhysRevLett.93.230801}.

\bibitem[{\citenamefont{Bear et~al.}(2000)\citenamefont{Bear, Stoner,
  Walsworth, Kosteleck{\'y}, and Lane}}]{Bear}
\bibinfo{author}{\bibfnamefont{D.}~\bibnamefont{Bear}},
  \bibinfo{author}{\bibfnamefont{R.~E.} \bibnamefont{Stoner}},
  \bibinfo{author}{\bibfnamefont{R.~L.} \bibnamefont{Walsworth}},
  \bibinfo{author}{\bibfnamefont{V.~A.} \bibnamefont{Kosteleck{\'y}}},
  \bibnamefont{and} \bibinfo{author}{\bibfnamefont{C.~D.} \bibnamefont{Lane}},
  \bibinfo{journal}{Physical Review Letters} \textbf{\bibinfo{volume}{85}},
  \bibinfo{pages}{5038} (\bibinfo{year}{2000}),
  \urlprefix\url{http://link.aps.org/doi/10.1103/PhysRevLett.85.5038}.

\bibitem[{\citenamefont{Hohensee et~al.}(2013)\citenamefont{Hohensee, Leefer,
  Budker, Harabati, Dzuba, and Flambaum}}]{Dy}
\bibinfo{author}{\bibfnamefont{M.~A.} \bibnamefont{Hohensee}},
  \bibinfo{author}{\bibfnamefont{N.}~\bibnamefont{Leefer}},
  \bibinfo{author}{\bibfnamefont{D.}~\bibnamefont{Budker}},
  \bibinfo{author}{\bibfnamefont{C.}~\bibnamefont{Harabati}},
  \bibinfo{author}{\bibfnamefont{V.~A.} \bibnamefont{Dzuba}}, \bibnamefont{and}
  \bibinfo{author}{\bibfnamefont{V.~V.} \bibnamefont{Flambaum}},
  \bibinfo{journal}{Physical Review Letters} \textbf{\bibinfo{volume}{111}},
  \bibinfo{pages}{050401} (\bibinfo{year}{2013}),
  \urlprefix\url{http://link.aps.org/doi/10.1103/PhysRevLett.111.050401}.

\bibitem[{\citenamefont{Peck et~al.}(2012)\citenamefont{Peck, Kim, Stein,
  Orbaker, Foss, Hummon, and Hunter}}]{Peck}
\bibinfo{author}{\bibfnamefont{S.~K.} \bibnamefont{Peck}},
  \bibinfo{author}{\bibfnamefont{D.~K.} \bibnamefont{Kim}},
  \bibinfo{author}{\bibfnamefont{D.}~\bibnamefont{Stein}},
  \bibinfo{author}{\bibfnamefont{D.}~\bibnamefont{Orbaker}},
  \bibinfo{author}{\bibfnamefont{A.}~\bibnamefont{Foss}},
  \bibinfo{author}{\bibfnamefont{M.~T.} \bibnamefont{Hummon}},
  \bibnamefont{and} \bibinfo{author}{\bibfnamefont{L.~R.}
  \bibnamefont{Hunter}}, \bibinfo{journal}{Physical Review A}
  \textbf{\bibinfo{volume}{86}}, \bibinfo{pages}{012109}
  (\bibinfo{year}{2012}),
  \urlprefix\url{http://link.aps.org/doi/10.1103/PhysRevA.86.012109}.

\bibitem[{\citenamefont{Brown et~al.}(2010)\citenamefont{Brown, Smullin,
  Kornack, and Romalis}}]{Brown}
\bibinfo{author}{\bibfnamefont{J.~M.} \bibnamefont{Brown}},
  \bibinfo{author}{\bibfnamefont{S.~J.} \bibnamefont{Smullin}},
  \bibinfo{author}{\bibfnamefont{T.~W.} \bibnamefont{Kornack}},
  \bibnamefont{and} \bibinfo{author}{\bibfnamefont{M.~V.}
  \bibnamefont{Romalis}}, \bibinfo{journal}{Physical Review Letters}
  \textbf{\bibinfo{volume}{105}}, \bibinfo{pages}{151604}
  (\bibinfo{year}{2010}),
  \urlprefix\url{http://link.aps.org/doi/10.1103/PhysRevLett.105.151604}.

\bibitem[{\citenamefont{Gemmel et~al.}(2010)\citenamefont{Gemmel, Heil, Karpuk,
  Lenz, Sobolev, Tullney, Burghoff, Kilian, Knappe-Gr{\"u}neberg, M{\"u}ller
  et~al.}}]{Gemmel}
\bibinfo{author}{\bibfnamefont{C.}~\bibnamefont{Gemmel}},
  \bibinfo{author}{\bibfnamefont{W.}~\bibnamefont{Heil}},
  \bibinfo{author}{\bibfnamefont{S.}~\bibnamefont{Karpuk}},
  \bibinfo{author}{\bibfnamefont{K.}~\bibnamefont{Lenz}},
  \bibinfo{author}{\bibfnamefont{Y.}~\bibnamefont{Sobolev}},
  \bibinfo{author}{\bibfnamefont{K.}~\bibnamefont{Tullney}},
  \bibinfo{author}{\bibfnamefont{M.}~\bibnamefont{Burghoff}},
  \bibinfo{author}{\bibfnamefont{W.}~\bibnamefont{Kilian}},
  \bibinfo{author}{\bibfnamefont{S.}~\bibnamefont{Knappe-Gr{\"u}neberg}},
  \bibinfo{author}{\bibfnamefont{W.}~\bibnamefont{M{\"u}ller}},
  \bibnamefont{et~al.}, \bibinfo{journal}{Physical Review D}
  \textbf{\bibinfo{volume}{82}}, \bibinfo{pages}{111901}
  (\bibinfo{year}{2010}),
  \urlprefix\url{http://link.aps.org/doi/10.1103/PhysRevD.82.111901}.

\bibitem[{\citenamefont{Altschul}(2009)}]{Altschul2009}
\bibinfo{author}{\bibfnamefont{B.}~\bibnamefont{Altschul}},
  \bibinfo{journal}{Physical Review D} \textbf{\bibinfo{volume}{79}},
  \bibinfo{pages}{061702} (\bibinfo{year}{2009}),
  \urlprefix\url{http://link.aps.org/doi/10.1103/PhysRevD.79.061702}.

\bibitem[{\citenamefont{Smiciklas et~al.}(2011)\citenamefont{Smiciklas, Brown,
  Cheuk, Smullin, and Romalis}}]{Smiciklas}
\bibinfo{author}{\bibfnamefont{M.}~\bibnamefont{Smiciklas}},
  \bibinfo{author}{\bibfnamefont{J.~M.} \bibnamefont{Brown}},
  \bibinfo{author}{\bibfnamefont{L.~W.} \bibnamefont{Cheuk}},
  \bibinfo{author}{\bibfnamefont{S.~J.} \bibnamefont{Smullin}},
  \bibnamefont{and} \bibinfo{author}{\bibfnamefont{M.~V.}
  \bibnamefont{Romalis}}, \bibinfo{journal}{Physical Review Letters}
  \textbf{\bibinfo{volume}{107}}, \bibinfo{pages}{171604}
  (\bibinfo{year}{2011}),
  \urlprefix\url{http://link.aps.org/doi/10.1103/PhysRevLett.107.171604}.

\bibitem[{\citenamefont{Altarev et~al.}(2009)\citenamefont{Altarev, Baker, Ban,
  Bison, Bodek, Daum, Fierlinger, Geltenbort, Green, van~der Grinten
  et~al.}}]{Altarev}
\bibinfo{author}{\bibfnamefont{I.}~\bibnamefont{Altarev}},
  \bibinfo{author}{\bibfnamefont{C.~A.} \bibnamefont{Baker}},
  \bibinfo{author}{\bibfnamefont{G.}~\bibnamefont{Ban}},
  \bibinfo{author}{\bibfnamefont{G.}~\bibnamefont{Bison}},
  \bibinfo{author}{\bibfnamefont{K.}~\bibnamefont{Bodek}},
  \bibinfo{author}{\bibfnamefont{M.}~\bibnamefont{Daum}},
  \bibinfo{author}{\bibfnamefont{P.}~\bibnamefont{Fierlinger}},
  \bibinfo{author}{\bibfnamefont{P.}~\bibnamefont{Geltenbort}},
  \bibinfo{author}{\bibfnamefont{K.}~\bibnamefont{Green}},
  \bibinfo{author}{\bibfnamefont{M.~G.~D.} \bibnamefont{van~der Grinten}},
  \bibnamefont{et~al.}, \bibinfo{journal}{Physical Review Letters}
  \textbf{\bibinfo{volume}{103}}, \bibinfo{pages}{081602}
  (\bibinfo{year}{2009}),
  \urlprefix\url{http://link.aps.org/doi/10.1103/PhysRevLett.103.081602}.

\bibitem[{\citenamefont{Pruttivarasin et~al.}(2015)\citenamefont{Pruttivarasin,
  Ramm, Porsev, Tupitsyn, Safronova, Hohensee, and Haffner}}]{Haeffner}
\bibinfo{author}{\bibfnamefont{T.}~\bibnamefont{Pruttivarasin}},
  \bibinfo{author}{\bibfnamefont{M.}~\bibnamefont{Ramm}},
  \bibinfo{author}{\bibfnamefont{S.~G.} \bibnamefont{Porsev}},
  \bibinfo{author}{\bibfnamefont{I.~I.} \bibnamefont{Tupitsyn}},
  \bibinfo{author}{\bibfnamefont{M.~S.} \bibnamefont{Safronova}},
  \bibinfo{author}{\bibfnamefont{M.~A.} \bibnamefont{Hohensee}},
  \bibnamefont{and} \bibinfo{author}{\bibfnamefont{H.}~\bibnamefont{Haffner}},
  \bibinfo{journal}{Nature} \textbf{\bibinfo{volume}{517}},
  \bibinfo{pages}{592} (\bibinfo{year}{2015}),
  \urlprefix\url{http://dx.doi.org/10.1038/nature14091}.

\bibitem[{\citenamefont{Heckel et~al.}(2008)\citenamefont{Heckel, Adelberger,
  Cramer, Cook, Schlamminger, and Schmidt}}]{Heckel}
\bibinfo{author}{\bibfnamefont{B.~R.} \bibnamefont{Heckel}},
  \bibinfo{author}{\bibfnamefont{E.~G.} \bibnamefont{Adelberger}},
  \bibinfo{author}{\bibfnamefont{C.~E.} \bibnamefont{Cramer}},
  \bibinfo{author}{\bibfnamefont{T.~S.} \bibnamefont{Cook}},
  \bibinfo{author}{\bibfnamefont{S.}~\bibnamefont{Schlamminger}},
  \bibnamefont{and} \bibinfo{author}{\bibfnamefont{U.}~\bibnamefont{Schmidt}},
  \bibinfo{journal}{Physical Review D} \textbf{\bibinfo{volume}{78}},
  \bibinfo{pages}{092006} (\bibinfo{year}{2008}),
  \urlprefix\url{http://link.aps.org/doi/10.1103/PhysRevD.78.092006}.

\bibitem[{\citenamefont{Bluhm et~al.}(2003)\citenamefont{Bluhm, Kosteleck{\'y},
  Lane, and Russell}}]{Bluhm}
\bibinfo{author}{\bibfnamefont{R.}~\bibnamefont{Bluhm}},
  \bibinfo{author}{\bibfnamefont{V.~A.} \bibnamefont{Kosteleck{\'y}}},
  \bibinfo{author}{\bibfnamefont{C.~D.} \bibnamefont{Lane}}, \bibnamefont{and}
  \bibinfo{author}{\bibfnamefont{N.}~\bibnamefont{Russell}},
  \bibinfo{journal}{Physical Review D} \textbf{\bibinfo{volume}{68}},
  \bibinfo{pages}{125008} (\bibinfo{year}{2003}),
  \urlprefix\url{http://link.aps.org/doi/10.1103/PhysRevD.68.125008}.

\bibitem[{\citenamefont{Gomes et~al.}(2014)\citenamefont{Gomes, Kosteleck{\'y},
  and Vargas}}]{Gomes}
\bibinfo{author}{\bibfnamefont{A.}~\bibnamefont{Gomes}},
  \bibinfo{author}{\bibfnamefont{V.~A.} \bibnamefont{Kosteleck{\'y}}},
  \bibnamefont{and} \bibinfo{author}{\bibfnamefont{A.~J.}
  \bibnamefont{Vargas}}, \bibinfo{journal}{Physical Review D}
  \textbf{\bibinfo{volume}{90}}, \bibinfo{pages}{076009}
  (\bibinfo{year}{2014}),
  \urlprefix\url{http://link.aps.org/doi/10.1103/PhysRevD.90.076009}.

\bibitem[{\citenamefont{Kosteleck{\'y} and Tasson}(2011)}]{KosteleckyTasson}
\bibinfo{author}{\bibfnamefont{V.~A.} \bibnamefont{Kosteleck{\'y}}}
  \bibnamefont{and} \bibinfo{author}{\bibfnamefont{J.~D.}
  \bibnamefont{Tasson}}, \bibinfo{journal}{Physical Review D}
  \textbf{\bibinfo{volume}{83}}, \bibinfo{pages}{016013}
  (\bibinfo{year}{2011}),
  \urlprefix\url{http://link.aps.org/doi/10.1103/PhysRevD.83.016013}.

\bibitem[{\citenamefont{Grudinin et~al.}(2010)\citenamefont{Grudinin, Lee,
  Painter, and Vahala}}]{phononlaser}
\bibinfo{author}{\bibfnamefont{I.~S.} \bibnamefont{Grudinin}},
  \bibinfo{author}{\bibfnamefont{H.}~\bibnamefont{Lee}},
  \bibinfo{author}{\bibfnamefont{O.}~\bibnamefont{Painter}}, \bibnamefont{and}
  \bibinfo{author}{\bibfnamefont{K.~J.} \bibnamefont{Vahala}},
  \bibinfo{journal}{Physical Review Letters} \textbf{\bibinfo{volume}{104}},
  \bibinfo{pages}{083901} (\bibinfo{year}{2010}),
  \urlprefix\url{http://link.aps.org/doi/10.1103/PhysRevLett.104.083901}.

\bibitem[{\citenamefont{Luan et~al.}(2014)\citenamefont{Luan, Huang, Li,
  McMillan, Zheng, Huang, Hsieh, Gu, Wang, Hati et~al.}}]{luan}
\bibinfo{author}{\bibfnamefont{X.}~\bibnamefont{Luan}},
  \bibinfo{author}{\bibfnamefont{Y.}~\bibnamefont{Huang}},
  \bibinfo{author}{\bibfnamefont{Y.}~\bibnamefont{Li}},
  \bibinfo{author}{\bibfnamefont{J.~F.} \bibnamefont{McMillan}},
  \bibinfo{author}{\bibfnamefont{J.}~\bibnamefont{Zheng}},
  \bibinfo{author}{\bibfnamefont{S.-W.} \bibnamefont{Huang}},
  \bibinfo{author}{\bibfnamefont{P.-C.} \bibnamefont{Hsieh}},
  \bibinfo{author}{\bibfnamefont{T.}~\bibnamefont{Gu}},
  \bibinfo{author}{\bibfnamefont{D.}~\bibnamefont{Wang}},
  \bibinfo{author}{\bibfnamefont{A.}~\bibnamefont{Hati}}, \bibnamefont{et~al.},
  \bibinfo{journal}{Scientific Reports} \textbf{\bibinfo{volume}{4}},
  \bibinfo{pages}{6842 EP } (\bibinfo{year}{2014}),
  \urlprefix\url{http://dx.doi.org/10.1038/srep06842}.

\bibitem[{\citenamefont{Seok et~al.}(2014)\citenamefont{Seok, Wright, and
  Meystre}}]{seok}
\bibinfo{author}{\bibfnamefont{H.}~\bibnamefont{Seok}},
  \bibinfo{author}{\bibfnamefont{E.~M.} \bibnamefont{Wright}},
  \bibnamefont{and} \bibinfo{author}{\bibfnamefont{P.}~\bibnamefont{Meystre}},
  \bibinfo{journal}{Physical Review A} \textbf{\bibinfo{volume}{90}},
  \bibinfo{pages}{043840} (\bibinfo{year}{2014}),
  \urlprefix\url{http://link.aps.org/doi/10.1103/PhysRevA.90.043840}.

\bibitem[{\citenamefont{Hossein-Zadeh and Vahala}(2008)}]{VahalaAPL}
\bibinfo{author}{\bibfnamefont{M.}~\bibnamefont{Hossein-Zadeh}}
  \bibnamefont{and} \bibinfo{author}{\bibfnamefont{K.~J.}
  \bibnamefont{Vahala}}, \bibinfo{journal}{Applied Physics Letters}
  \textbf{\bibinfo{volume}{93}}, \bibinfo{pages}{191115}
  (\bibinfo{year}{2008}),
  \urlprefix\url{http://scitation.aip.org/content/aip/journal/apl/93/19/10.1063/1.3028024}.

\bibitem[{\citenamefont{Li et~al.}(2014)\citenamefont{Li, Yi, Lee, Diddams, and
  Vahala}}]{VahalaScience}
\bibinfo{author}{\bibfnamefont{J.}~\bibnamefont{Li}},
  \bibinfo{author}{\bibfnamefont{X.}~\bibnamefont{Yi}},
  \bibinfo{author}{\bibfnamefont{H.}~\bibnamefont{Lee}},
  \bibinfo{author}{\bibfnamefont{S.~A.} \bibnamefont{Diddams}},
  \bibnamefont{and} \bibinfo{author}{\bibfnamefont{K.~J.}
  \bibnamefont{Vahala}}, \bibinfo{journal}{Science}
  \textbf{\bibinfo{volume}{345}}, \bibinfo{pages}{309} (\bibinfo{year}{2014}),
  \urlprefix\url{http://www.sciencemag.org/content/345/6194/309.abstract}.

\bibitem[{\citenamefont{Tomes and Carmon}(2009)}]{Tomes}
\bibinfo{author}{\bibfnamefont{M.}~\bibnamefont{Tomes}} \bibnamefont{and}
  \bibinfo{author}{\bibfnamefont{T.}~\bibnamefont{Carmon}},
  \bibinfo{journal}{Physical Review Letters} \textbf{\bibinfo{volume}{102}},
  \bibinfo{pages}{113601} (\bibinfo{year}{2009}),
  \urlprefix\url{http://link.aps.org/doi/10.1103/PhysRevLett.102.113601}.

\bibitem[{\citenamefont{Grudinin et~al.}(2009)\citenamefont{Grudinin, Matsko,
  and Maleki}}]{Grudinin}
\bibinfo{author}{\bibfnamefont{I.~S.} \bibnamefont{Grudinin}},
  \bibinfo{author}{\bibfnamefont{A.~B.} \bibnamefont{Matsko}},
  \bibnamefont{and} \bibinfo{author}{\bibfnamefont{L.}~\bibnamefont{Maleki}},
  \bibinfo{journal}{Physical Review Letters} \textbf{\bibinfo{volume}{102}},
  \bibinfo{pages}{043902} (\bibinfo{year}{2009}),
  \urlprefix\url{http://link.aps.org/doi/10.1103/PhysRevLett.102.043902}.

\bibitem[{\citenamefont{Lee et~al.}(2012)\citenamefont{Lee, Chen, Li, Yang,
  Jeon, Painter, and Vahala}}]{Lee}
\bibinfo{author}{\bibfnamefont{H.}~\bibnamefont{Lee}},
  \bibinfo{author}{\bibfnamefont{T.}~\bibnamefont{Chen}},
  \bibinfo{author}{\bibfnamefont{J.}~\bibnamefont{Li}},
  \bibinfo{author}{\bibfnamefont{K.~Y.} \bibnamefont{Yang}},
  \bibinfo{author}{\bibfnamefont{S.}~\bibnamefont{Jeon}},
  \bibinfo{author}{\bibfnamefont{O.}~\bibnamefont{Painter}}, \bibnamefont{and}
  \bibinfo{author}{\bibfnamefont{K.~J.} \bibnamefont{Vahala}},
  \bibinfo{journal}{Nat Photon} \textbf{\bibinfo{volume}{6}},
  \bibinfo{pages}{369} (\bibinfo{year}{2012}),
  \urlprefix\url{http://dx.doi.org/10.1038/nphoton.2012.109}.

\bibitem[{\citenamefont{Li et~al.}(2012)\citenamefont{Li, Lee, Chen, and
  Vahala}}]{Li}
\bibinfo{author}{\bibfnamefont{J.}~\bibnamefont{Li}},
  \bibinfo{author}{\bibfnamefont{H.}~\bibnamefont{Lee}},
  \bibinfo{author}{\bibfnamefont{T.}~\bibnamefont{Chen}}, \bibnamefont{and}
  \bibinfo{author}{\bibfnamefont{K.~J.} \bibnamefont{Vahala}},
  \bibinfo{journal}{Optics Express} \textbf{\bibinfo{volume}{20}},
  \bibinfo{pages}{20170} (\bibinfo{year}{2012}),
  \urlprefix\url{http://www.opticsexpress.org/abstract.cfm?URI=oe-20-18-20170}.

\bibitem[{\citenamefont{Kosteleck{\'y} and Russell}(2011)}]{datatables}
\bibinfo{author}{\bibfnamefont{V.~A.} \bibnamefont{Kosteleck{\'y}}}
  \bibnamefont{and} \bibinfo{author}{\bibfnamefont{N.}~\bibnamefont{Russell}},
  \bibinfo{journal}{Reviews of Modern Physics, arXiv:0801.0287}
  \textbf{\bibinfo{volume}{83}}, \bibinfo{pages}{11} (\bibinfo{year}{2011}),
  \urlprefix\url{http://link.aps.org/doi/10.1103/RevModPhys.83.11}.

\bibitem[{\citenamefont{Lightman and Lee}(1973)}]{Lightman1973}
\bibinfo{author}{\bibfnamefont{A.~P.} \bibnamefont{Lightman}} \bibnamefont{and}
  \bibinfo{author}{\bibfnamefont{D.~L.} \bibnamefont{Lee}},
  \bibinfo{journal}{Physical Review D} \textbf{\bibinfo{volume}{8}},
  \bibinfo{pages}{364} (\bibinfo{year}{1973}),
  \urlprefix\url{http://link.aps.org/doi/10.1103/PhysRevD.8.364}.

\bibitem[{\citenamefont{Nielsen and Picek}(1983)}]{Nielsen1983}
\bibinfo{author}{\bibfnamefont{H.~B.} \bibnamefont{Nielsen}} \bibnamefont{and}
  \bibinfo{author}{\bibfnamefont{I.}~\bibnamefont{Picek}},
  \bibinfo{journal}{Nuclear Physics B} \textbf{\bibinfo{volume}{211}},
  \bibinfo{pages}{269} (\bibinfo{year}{1983}),
  \urlprefix\url{http://www.sciencedirect.com/science/article/pii/0550321383904091}.

\bibitem[{\citenamefont{Will}(1993)}]{Will1993}
\bibinfo{author}{\bibfnamefont{C.}~\bibnamefont{Will}},
  \emph{\bibinfo{title}{Theory and Experiment in Gravitational Physics}}
  (\bibinfo{publisher}{Cambridge University Press},
  \bibinfo{address}{Cambridge}, \bibinfo{year}{1993}).

\bibitem[{\citenamefont{Coleman and Glashow}(1999)}]{Coleman1999}
\bibinfo{author}{\bibfnamefont{S.}~\bibnamefont{Coleman}} \bibnamefont{and}
  \bibinfo{author}{\bibfnamefont{S.~L.} \bibnamefont{Glashow}},
  \bibinfo{journal}{Physical Review D} \textbf{\bibinfo{volume}{59}},
  \bibinfo{pages}{116008} (\bibinfo{year}{1999}),
  \urlprefix\url{http://link.aps.org/doi/10.1103/PhysRevD.59.116008}.

\bibitem[{\citenamefont{Kosteleck{\'y} and Mewes}(2013)}]{Mewes}
\bibinfo{author}{\bibfnamefont{V.~A.} \bibnamefont{Kosteleck{\'y}}}
  \bibnamefont{and} \bibinfo{author}{\bibfnamefont{M.}~\bibnamefont{Mewes}},
  \bibinfo{journal}{Physical Review Letters} \textbf{\bibinfo{volume}{110}},
  \bibinfo{pages}{201601} (\bibinfo{year}{2013}),
  \urlprefix\url{http://link.aps.org/doi/10.1103/PhysRevLett.110.201601}.

\bibitem[{\citenamefont{Michelson and Morley}(1887{\natexlab{a}})}]{MM1}
\bibinfo{author}{\bibfnamefont{A.}~\bibnamefont{Michelson}} \bibnamefont{and}
  \bibinfo{author}{\bibfnamefont{E.}~\bibnamefont{Morley}},
  \bibinfo{journal}{Am. J. Sci.} \textbf{\bibinfo{volume}{34}},
  \bibinfo{pages}{333} (\bibinfo{year}{1887}{\natexlab{a}}).

\bibitem[{\citenamefont{Michelson and Morley}(1887{\natexlab{b}})}]{MM2}
\bibinfo{author}{\bibfnamefont{A.}~\bibnamefont{Michelson}} \bibnamefont{and}
  \bibinfo{author}{\bibfnamefont{E.}~\bibnamefont{Morley}},
  \bibinfo{journal}{Phil. Mag.} \textbf{\bibinfo{volume}{24}},
  \bibinfo{pages}{449} (\bibinfo{year}{1887}{\natexlab{b}}).

\bibitem[{\citenamefont{Kennedy and Thorndike}(1932)}]{MM3}
\bibinfo{author}{\bibfnamefont{R.}~\bibnamefont{Kennedy}} \bibnamefont{and}
  \bibinfo{author}{\bibfnamefont{E.}~\bibnamefont{Thorndike}},
  \bibinfo{journal}{Phys. Rev.} \textbf{\bibinfo{volume}{42}},
  \bibinfo{pages}{400} (\bibinfo{year}{1932}).

\bibitem[{\citenamefont{Ives and Stilwell}(1938)}]{MM4}
\bibinfo{author}{\bibfnamefont{H.~E.} \bibnamefont{Ives}} \bibnamefont{and}
  \bibinfo{author}{\bibfnamefont{G.~R.} \bibnamefont{Stilwell}},
  \bibinfo{journal}{Journal of the Optical Society of America}
  \textbf{\bibinfo{volume}{28}}, \bibinfo{pages}{215} (\bibinfo{year}{1938}),
  \urlprefix\url{http://www.osapublishing.org/abstract.cfm?URI=josa-28-7-215}.

\bibitem[{\citenamefont{Hohensee et~al.}(2007)\citenamefont{Hohensee, Glenday,
  Li, Tobar, and Wolf}}]{Photons1}
\bibinfo{author}{\bibfnamefont{M.}~\bibnamefont{Hohensee}},
  \bibinfo{author}{\bibfnamefont{A.}~\bibnamefont{Glenday}},
  \bibinfo{author}{\bibfnamefont{C.-H.} \bibnamefont{Li}},
  \bibinfo{author}{\bibfnamefont{M.~E.} \bibnamefont{Tobar}}, \bibnamefont{and}
  \bibinfo{author}{\bibfnamefont{P.}~\bibnamefont{Wolf}},
  \bibinfo{journal}{Physical Review D} \textbf{\bibinfo{volume}{75}},
  \bibinfo{pages}{049902} (\bibinfo{year}{2007}),
  \urlprefix\url{http://link.aps.org/doi/10.1103/PhysRevD.75.049902}.

\bibitem[{\citenamefont{Stanwix et~al.}(2006)\citenamefont{Stanwix, Tobar,
  Wolf, Locke, and Ivanov}}]{Photons2}
\bibinfo{author}{\bibfnamefont{P.~L.} \bibnamefont{Stanwix}},
  \bibinfo{author}{\bibfnamefont{M.~E.} \bibnamefont{Tobar}},
  \bibinfo{author}{\bibfnamefont{P.}~\bibnamefont{Wolf}},
  \bibinfo{author}{\bibfnamefont{C.~R.} \bibnamefont{Locke}}, \bibnamefont{and}
  \bibinfo{author}{\bibfnamefont{E.~N.} \bibnamefont{Ivanov}},
  \bibinfo{journal}{Physical Review D} \textbf{\bibinfo{volume}{74}},
  \bibinfo{pages}{081101} (\bibinfo{year}{2006}),
  \urlprefix\url{http://link.aps.org/doi/10.1103/PhysRevD.74.081101}.

\bibitem[{\citenamefont{Antonini
  et~al.}(2005{\natexlab{a}})\citenamefont{Antonini, Okhapkin, G{\"o}kl{\"u},
  and Schiller}}]{Photons3}
\bibinfo{author}{\bibfnamefont{P.}~\bibnamefont{Antonini}},
  \bibinfo{author}{\bibfnamefont{M.}~\bibnamefont{Okhapkin}},
  \bibinfo{author}{\bibfnamefont{E.}~\bibnamefont{G{\"o}kl{\"u}}},
  \bibnamefont{and} \bibinfo{author}{\bibfnamefont{S.}~\bibnamefont{Schiller}},
  \bibinfo{journal}{Physical Review A} \textbf{\bibinfo{volume}{72}},
  \bibinfo{pages}{066102} (\bibinfo{year}{2005}{\natexlab{a}}),
  \urlprefix\url{http://link.aps.org/doi/10.1103/PhysRevA.72.066102}.

\bibitem[{\citenamefont{Tobar et~al.}(2005{\natexlab{a}})\citenamefont{Tobar,
  Wolf, and Stanwix}}]{Photons4}
\bibinfo{author}{\bibfnamefont{M.~E.} \bibnamefont{Tobar}},
  \bibinfo{author}{\bibfnamefont{P.}~\bibnamefont{Wolf}}, \bibnamefont{and}
  \bibinfo{author}{\bibfnamefont{P.~L.} \bibnamefont{Stanwix}},
  \bibinfo{journal}{Physical Review A} \textbf{\bibinfo{volume}{72}},
  \bibinfo{pages}{066101} (\bibinfo{year}{2005}{\natexlab{a}}),
  \urlprefix\url{http://link.aps.org/doi/10.1103/PhysRevA.72.066101}.

\bibitem[{\citenamefont{Herrmann et~al.}(2005)\citenamefont{Herrmann, Senger,
  Kovalchuk, M{\"u}ller, and Peters}}]{Photons5}
\bibinfo{author}{\bibfnamefont{S.}~\bibnamefont{Herrmann}},
  \bibinfo{author}{\bibfnamefont{A.}~\bibnamefont{Senger}},
  \bibinfo{author}{\bibfnamefont{E.}~\bibnamefont{Kovalchuk}},
  \bibinfo{author}{\bibfnamefont{H.}~\bibnamefont{M{\"u}ller}},
  \bibnamefont{and} \bibinfo{author}{\bibfnamefont{A.}~\bibnamefont{Peters}},
  \bibinfo{journal}{Physical Review Letters} \textbf{\bibinfo{volume}{95}},
  \bibinfo{pages}{150401} (\bibinfo{year}{2005}),
  \urlprefix\url{http://link.aps.org/doi/10.1103/PhysRevLett.95.150401}.

\bibitem[{\citenamefont{Ehlers et~al.}(2006)\citenamefont{Ehlers,
  L{\"a}mmerzahl, Tobar, Stanwix, Susli, Wolf, Locke, and Ivanov}}]{Photons6}
\bibinfo{author}{\bibfnamefont{J.}~\bibnamefont{Ehlers}},
  \bibinfo{author}{\bibfnamefont{C.}~\bibnamefont{L{\"a}mmerzahl}},
  \bibinfo{author}{\bibfnamefont{M.~E.} \bibnamefont{Tobar}},
  \bibinfo{author}{\bibfnamefont{P.~L.} \bibnamefont{Stanwix}},
  \bibinfo{author}{\bibfnamefont{M.}~\bibnamefont{Susli}},
  \bibinfo{author}{\bibfnamefont{P.}~\bibnamefont{Wolf}},
  \bibinfo{author}{\bibfnamefont{C.~R.} \bibnamefont{Locke}}, \bibnamefont{and}
  \bibinfo{author}{\bibfnamefont{E.~N.} \bibnamefont{Ivanov}},
  \emph{\bibinfo{title}{Lecture Notes in Physics}}
  (\bibinfo{publisher}{Springer Berlin Heidelberg}, \bibinfo{year}{2006}), vol.
  \bibinfo{volume}{702}, pp. \bibinfo{pages}{416--450}, ISBN
  \bibinfo{isbn}{978-3-540-34522-0},
  \urlprefix\url{http://dx.doi.org/10.1007/3-540-34523-X_15}.

\bibitem[{\citenamefont{Stanwix
  et~al.}(2005{\natexlab{b}})\citenamefont{Stanwix, Tobar, Wolf, Susli, Locke,
  Ivanov, Winterflood, and van Kann}}]{Photons7}
\bibinfo{author}{\bibfnamefont{P.~L.} \bibnamefont{Stanwix}},
  \bibinfo{author}{\bibfnamefont{M.~E.} \bibnamefont{Tobar}},
  \bibinfo{author}{\bibfnamefont{P.}~\bibnamefont{Wolf}},
  \bibinfo{author}{\bibfnamefont{M.}~\bibnamefont{Susli}},
  \bibinfo{author}{\bibfnamefont{C.~R.} \bibnamefont{Locke}},
  \bibinfo{author}{\bibfnamefont{E.~N.} \bibnamefont{Ivanov}},
  \bibinfo{author}{\bibfnamefont{J.}~\bibnamefont{Winterflood}},
  \bibnamefont{and} \bibinfo{author}{\bibfnamefont{F.}~\bibnamefont{van Kann}},
  \bibinfo{journal}{Physical Review Letters} \textbf{\bibinfo{volume}{95}},
  \bibinfo{pages}{040404} (\bibinfo{year}{2005}{\natexlab{b}}),
  \urlprefix\url{http://link.aps.org/doi/10.1103/PhysRevLett.95.040404}.

\bibitem[{\citenamefont{Antonini
  et~al.}(2005{\natexlab{b}})\citenamefont{Antonini, Okhapkin, G{\"o}kl{\"u},
  and Schiller}}]{Photons8}
\bibinfo{author}{\bibfnamefont{P.}~\bibnamefont{Antonini}},
  \bibinfo{author}{\bibfnamefont{M.}~\bibnamefont{Okhapkin}},
  \bibinfo{author}{\bibfnamefont{E.}~\bibnamefont{G{\"o}kl{\"u}}},
  \bibnamefont{and} \bibinfo{author}{\bibfnamefont{S.}~\bibnamefont{Schiller}},
  \bibinfo{journal}{Physical Review A} \textbf{\bibinfo{volume}{71}},
  \bibinfo{pages}{050101} (\bibinfo{year}{2005}{\natexlab{b}}),
  \urlprefix\url{http://link.aps.org/doi/10.1103/PhysRevA.71.050101}.

\bibitem[{\citenamefont{Tobar et~al.}(2005{\natexlab{b}})\citenamefont{Tobar,
  Wolf, Fowler, and Hartnett}}]{Photons9}
\bibinfo{author}{\bibfnamefont{M.~E.} \bibnamefont{Tobar}},
  \bibinfo{author}{\bibfnamefont{P.}~\bibnamefont{Wolf}},
  \bibinfo{author}{\bibfnamefont{A.}~\bibnamefont{Fowler}}, \bibnamefont{and}
  \bibinfo{author}{\bibfnamefont{J.~G.} \bibnamefont{Hartnett}},
  \bibinfo{journal}{Physical Review D} \textbf{\bibinfo{volume}{71}},
  \bibinfo{pages}{025004} (\bibinfo{year}{2005}{\natexlab{b}}),
  \urlprefix\url{http://link.aps.org/doi/10.1103/PhysRevD.71.025004}.

\bibitem[{\citenamefont{Wolf et~al.}(2004)\citenamefont{Wolf, Bize, Clairon,
  Santarelli, Tobar, and Luiten}}]{Photons10}
\bibinfo{author}{\bibfnamefont{P.}~\bibnamefont{Wolf}},
  \bibinfo{author}{\bibfnamefont{S.}~\bibnamefont{Bize}},
  \bibinfo{author}{\bibfnamefont{A.}~\bibnamefont{Clairon}},
  \bibinfo{author}{\bibfnamefont{G.}~\bibnamefont{Santarelli}},
  \bibinfo{author}{\bibfnamefont{M.~E.} \bibnamefont{Tobar}}, \bibnamefont{and}
  \bibinfo{author}{\bibfnamefont{A.}~\bibnamefont{Luiten}},
  \bibinfo{journal}{Physical Review D} \textbf{\bibinfo{volume}{70}},
  \bibinfo{pages}{051902} (\bibinfo{year}{2004}),
  \urlprefix\url{http://link.aps.org/doi/10.1103/PhysRevD.70.051902}.

\bibitem[{\citenamefont{M{\"u}ller
  et~al.}(2003{\natexlab{a}})\citenamefont{M{\"u}ller, Herrmann, Saenz, Peters,
  and L{\"a}mmerzahl}}]{Photons11}
\bibinfo{author}{\bibfnamefont{H.}~\bibnamefont{M{\"u}ller}},
  \bibinfo{author}{\bibfnamefont{S.}~\bibnamefont{Herrmann}},
  \bibinfo{author}{\bibfnamefont{A.}~\bibnamefont{Saenz}},
  \bibinfo{author}{\bibfnamefont{A.}~\bibnamefont{Peters}}, \bibnamefont{and}
  \bibinfo{author}{\bibfnamefont{C.}~\bibnamefont{L{\"a}mmerzahl}},
  \bibinfo{journal}{Physical Review D} \textbf{\bibinfo{volume}{68}},
  \bibinfo{pages}{116006} (\bibinfo{year}{2003}{\natexlab{a}}),
  \urlprefix\url{http://link.aps.org/doi/10.1103/PhysRevD.68.116006}.

\bibitem[{\citenamefont{M{\"u}ller
  et~al.}(2003{\natexlab{b}})\citenamefont{M{\"u}ller, Herrmann, Braxmaier,
  Schiller, and Peters}}]{Photons12}
\bibinfo{author}{\bibfnamefont{H.}~\bibnamefont{M{\"u}ller}},
  \bibinfo{author}{\bibfnamefont{S.}~\bibnamefont{Herrmann}},
  \bibinfo{author}{\bibfnamefont{C.}~\bibnamefont{Braxmaier}},
  \bibinfo{author}{\bibfnamefont{S.}~\bibnamefont{Schiller}}, \bibnamefont{and}
  \bibinfo{author}{\bibfnamefont{A.}~\bibnamefont{Peters}},
  \bibinfo{journal}{Physical Review Letters} \textbf{\bibinfo{volume}{91}},
  \bibinfo{pages}{020401} (\bibinfo{year}{2003}{\natexlab{b}}),
  \urlprefix\url{http://link.aps.org/doi/10.1103/PhysRevLett.91.020401}.

\bibitem[{\citenamefont{Lipa et~al.}(2003)\citenamefont{Lipa, Nissen, Wang,
  Stricker, and Avaloff}}]{Photons13}
\bibinfo{author}{\bibfnamefont{J.~A.} \bibnamefont{Lipa}},
  \bibinfo{author}{\bibfnamefont{J.~A.} \bibnamefont{Nissen}},
  \bibinfo{author}{\bibfnamefont{S.}~\bibnamefont{Wang}},
  \bibinfo{author}{\bibfnamefont{D.~A.} \bibnamefont{Stricker}},
  \bibnamefont{and} \bibinfo{author}{\bibfnamefont{D.}~\bibnamefont{Avaloff}},
  \bibinfo{journal}{Physical Review Letters} \textbf{\bibinfo{volume}{90}},
  \bibinfo{pages}{060403} (\bibinfo{year}{2003}),
  \urlprefix\url{http://link.aps.org/doi/10.1103/PhysRevLett.90.060403}.

\bibitem[{\citenamefont{Nagel et~al.}(2015)\citenamefont{Nagel, Parker,
  Kovalchuk, Stanwix, Hartnett, Ivanov, Peters, and Tobar}}]{Nagel}
\bibinfo{author}{\bibfnamefont{M.}~\bibnamefont{Nagel}},
  \bibinfo{author}{\bibfnamefont{S.~R.} \bibnamefont{Parker}},
  \bibinfo{author}{\bibfnamefont{E.~V.} \bibnamefont{Kovalchuk}},
  \bibinfo{author}{\bibfnamefont{P.~L.} \bibnamefont{Stanwix}},
  \bibinfo{author}{\bibfnamefont{J.~G.} \bibnamefont{Hartnett}},
  \bibinfo{author}{\bibfnamefont{E.~N.} \bibnamefont{Ivanov}},
  \bibinfo{author}{\bibfnamefont{A.}~\bibnamefont{Peters}}, \bibnamefont{and}
  \bibinfo{author}{\bibfnamefont{M.~E.} \bibnamefont{Tobar}},
  \bibinfo{journal}{Nat Commun} \textbf{\bibinfo{volume}{6}}
  (\bibinfo{year}{2015}), \urlprefix\url{http://dx.doi.org/10.1038/ncomms9174}.

\bibitem[{\citenamefont{Hohensee et~al.}(2010)\citenamefont{Hohensee, Stanwix,
  Tobar, Parker, Phillips, and Walsworth}}]{Hohensee}
\bibinfo{author}{\bibfnamefont{M.~A.} \bibnamefont{Hohensee}},
  \bibinfo{author}{\bibfnamefont{P.~L.} \bibnamefont{Stanwix}},
  \bibinfo{author}{\bibfnamefont{M.~E.} \bibnamefont{Tobar}},
  \bibinfo{author}{\bibfnamefont{S.~R.} \bibnamefont{Parker}},
  \bibinfo{author}{\bibfnamefont{D.~F.} \bibnamefont{Phillips}},
  \bibnamefont{and} \bibinfo{author}{\bibfnamefont{R.~L.}
  \bibnamefont{Walsworth}}, \bibinfo{journal}{Physical Review D}
  \textbf{\bibinfo{volume}{82}}, \bibinfo{pages}{076001}
  (\bibinfo{year}{2010}),
  \urlprefix\url{http://link.aps.org/doi/10.1103/PhysRevD.82.076001}.

\bibitem[{\citenamefont{Baynes et~al.}(2012)\citenamefont{Baynes, Tobar, and
  Luiten}}]{Baynes}
\bibinfo{author}{\bibfnamefont{F.~N.} \bibnamefont{Baynes}},
  \bibinfo{author}{\bibfnamefont{M.~E.} \bibnamefont{Tobar}}, \bibnamefont{and}
  \bibinfo{author}{\bibfnamefont{A.~N.} \bibnamefont{Luiten}},
  \bibinfo{journal}{Physical Review Letters} \textbf{\bibinfo{volume}{108}},
  \bibinfo{pages}{260801} (\bibinfo{year}{2012}),
  \urlprefix\url{http://link.aps.org/doi/10.1103/PhysRevLett.108.260801}.

\bibitem[{pz:(1988)}]{pz:1988zr}
\bibinfo{journal}{ANSI/IEEE Std 176-1987} pp. \bibinfo{pages}{0{\_}1--}
  (\bibinfo{year}{1988}).

\bibitem[{\citenamefont{Filler}(May 1988)}]{Filler:1988oa}
\bibinfo{author}{\bibfnamefont{R.~L.} \bibnamefont{Filler}},
  \bibinfo{journal}{Ultrasonics, Ferroelectrics and Frequency Control, IEEE
  Transactions on} \textbf{\bibinfo{volume}{35}}, \bibinfo{pages}{297}
  (\bibinfo{year}{May 1988}).

\bibitem[{\citenamefont{Stevens and Tiersten}(1986)}]{Stevens}
\bibinfo{author}{\bibfnamefont{D.~S.} \bibnamefont{Stevens}} \bibnamefont{and}
  \bibinfo{author}{\bibfnamefont{H.~F.} \bibnamefont{Tiersten}},
  \bibinfo{journal}{The Journal of the Acoustical Society of America}
  \textbf{\bibinfo{volume}{79}}, \bibinfo{pages}{1811} (\bibinfo{year}{1986}).

\bibitem[{\citenamefont{EerNisse}(2001)}]{Eernisse}
\bibinfo{author}{\bibfnamefont{E.}~\bibnamefont{EerNisse}},
  \bibinfo{journal}{IEEE UFFC} \textbf{\bibinfo{volume}{48}},
  \bibinfo{pages}{1351} (\bibinfo{year}{2001}).

\bibitem[{\citenamefont{Prestage et~al.}(1985)\citenamefont{Prestage,
  Bollinger, Itano, and Wineland}}]{Prestage}
\bibinfo{author}{\bibfnamefont{J.~D.} \bibnamefont{Prestage}},
  \bibinfo{author}{\bibfnamefont{J.~J.} \bibnamefont{Bollinger}},
  \bibinfo{author}{\bibfnamefont{W.~M.} \bibnamefont{Itano}}, \bibnamefont{and}
  \bibinfo{author}{\bibfnamefont{D.~J.} \bibnamefont{Wineland}},
  \bibinfo{journal}{Physical Review Letters} \textbf{\bibinfo{volume}{54}},
  \bibinfo{pages}{2387} (\bibinfo{year}{1985}),
  \urlprefix\url{http://link.aps.org/doi/10.1103/PhysRevLett.54.2387}.

\bibitem[{\citenamefont{Lamoreaux et~al.}(1986)\citenamefont{Lamoreaux, Jacobs,
  Heckel, Raab, and Fortson}}]{Lamoreaux}
\bibinfo{author}{\bibfnamefont{S.~K.} \bibnamefont{Lamoreaux}},
  \bibinfo{author}{\bibfnamefont{J.~P.} \bibnamefont{Jacobs}},
  \bibinfo{author}{\bibfnamefont{B.~R.} \bibnamefont{Heckel}},
  \bibinfo{author}{\bibfnamefont{F.~J.} \bibnamefont{Raab}}, \bibnamefont{and}
  \bibinfo{author}{\bibfnamefont{E.~N.} \bibnamefont{Fortson}},
  \bibinfo{journal}{Phys. Rev. Lett.} \textbf{\bibinfo{volume}{57}},
  \bibinfo{pages}{3125} (\bibinfo{year}{1986}),
  \urlprefix\url{http://link.aps.org/doi/10.1103/PhysRevLett.57.3125}.

\bibitem[{\citenamefont{Chupp et~al.}(1989)\citenamefont{Chupp, Hoare, Loveman,
  Oteiza, Richardson, Wagshul, and Thompson}}]{Chupp}
\bibinfo{author}{\bibfnamefont{T.~E.} \bibnamefont{Chupp}},
  \bibinfo{author}{\bibfnamefont{R.~J.} \bibnamefont{Hoare}},
  \bibinfo{author}{\bibfnamefont{R.~A.} \bibnamefont{Loveman}},
  \bibinfo{author}{\bibfnamefont{E.~R.} \bibnamefont{Oteiza}},
  \bibinfo{author}{\bibfnamefont{J.~M.} \bibnamefont{Richardson}},
  \bibinfo{author}{\bibfnamefont{M.~E.} \bibnamefont{Wagshul}},
  \bibnamefont{and} \bibinfo{author}{\bibfnamefont{A.~K.}
  \bibnamefont{Thompson}}, \bibinfo{journal}{Phys. Rev. Lett.}
  \textbf{\bibinfo{volume}{63}}, \bibinfo{pages}{1541} (\bibinfo{year}{1989}),
  \urlprefix\url{http://link.aps.org/doi/10.1103/PhysRevLett.63.1541}.

\bibitem[{\citenamefont{Altschul}(2008)}]{Altschul}
\bibinfo{author}{\bibfnamefont{B.}~\bibnamefont{Altschul}},
  \bibinfo{journal}{Physical Review D} \textbf{\bibinfo{volume}{78}},
  \bibinfo{pages}{085018} (\bibinfo{year}{2008}),
  \urlprefix\url{http://link.aps.org/doi/10.1103/PhysRevD.78.085018}.

\bibitem[{\citenamefont{Hohensee et~al.}(2011)\citenamefont{Hohensee, Chu,
  Peters, and M{\"u}ller}}]{redshiftPRL}
\bibinfo{author}{\bibfnamefont{M.~A.} \bibnamefont{Hohensee}},
  \bibinfo{author}{\bibfnamefont{S.}~\bibnamefont{Chu}},
  \bibinfo{author}{\bibfnamefont{A.}~\bibnamefont{Peters}}, \bibnamefont{and}
  \bibinfo{author}{\bibfnamefont{H.}~\bibnamefont{M{\"u}ller}},
  \bibinfo{journal}{Physical Review Letters} \textbf{\bibinfo{volume}{106}},
  \bibinfo{pages}{151102} (\bibinfo{year}{2011}),
  \urlprefix\url{http://link.aps.org/doi/10.1103/PhysRevLett.106.151102}.

\bibitem[{\citenamefont{Salzenstein et~al.}(October
  2010)\citenamefont{Salzenstein, Kuna, Sojdr, and
  Chauvin}}]{Salzenstein:2010aa}
\bibinfo{author}{\bibfnamefont{P.}~\bibnamefont{Salzenstein}},
  \bibinfo{author}{\bibfnamefont{A.}~\bibnamefont{Kuna}},
  \bibinfo{author}{\bibfnamefont{L.}~\bibnamefont{Sojdr}}, \bibnamefont{and}
  \bibinfo{author}{\bibfnamefont{J.}~\bibnamefont{Chauvin}},
  \bibinfo{journal}{Electronics Letters} \textbf{\bibinfo{volume}{46}},
  \bibinfo{pages}{1433} (\bibinfo{year}{October 2010}).

\bibitem[{\citenamefont{Besson}(1977)}]{1537081}
\bibinfo{author}{\bibfnamefont{R.~J.} \bibnamefont{Besson}}, in
  \emph{\bibinfo{booktitle}{31st Annual Symposium on Frequency Control}}
  (\bibinfo{year}{1977}), pp. \bibinfo{pages}{147 -- 152}.

\bibitem[{\citenamefont{Galliou et~al.}(2011)\citenamefont{Galliou, Imbaud,
  Goryachev, Bourquin, and Abbe}}]{galliou:091911}
\bibinfo{author}{\bibfnamefont{S.}~\bibnamefont{Galliou}},
  \bibinfo{author}{\bibfnamefont{J.}~\bibnamefont{Imbaud}},
  \bibinfo{author}{\bibfnamefont{M.}~\bibnamefont{Goryachev}},
  \bibinfo{author}{\bibfnamefont{R.}~\bibnamefont{Bourquin}}, \bibnamefont{and}
  \bibinfo{author}{\bibfnamefont{P.}~\bibnamefont{Abbe}},
  \bibinfo{journal}{Applied Physics Letters} \textbf{\bibinfo{volume}{98}},
  \bibinfo{pages}{091911} (\bibinfo{year}{2011}).

\bibitem[{\citenamefont{Goryachev et~al.}(2012)\citenamefont{Goryachev,
  Creedon, Ivanov, Galliou, Bourquin, and Tobar}}]{ours}
\bibinfo{author}{\bibfnamefont{M.}~\bibnamefont{Goryachev}},
  \bibinfo{author}{\bibfnamefont{D.}~\bibnamefont{Creedon}},
  \bibinfo{author}{\bibfnamefont{E.}~\bibnamefont{Ivanov}},
  \bibinfo{author}{\bibfnamefont{S.}~\bibnamefont{Galliou}},
  \bibinfo{author}{\bibfnamefont{R.}~\bibnamefont{Bourquin}}, \bibnamefont{and}
  \bibinfo{author}{\bibfnamefont{M.}~\bibnamefont{Tobar}},
  \bibinfo{journal}{Applied Physics Letters} \textbf{\bibinfo{volume}{100}},
  \bibinfo{pages}{243504} (\bibinfo{year}{2012}).

\bibitem[{\citenamefont{Goryachev
  et~al.}(2013{\natexlab{a}})\citenamefont{Goryachev, Creedon, Galliou, and
  Tobar}}]{quartzPRL}
\bibinfo{author}{\bibfnamefont{M.}~\bibnamefont{Goryachev}},
  \bibinfo{author}{\bibfnamefont{D.}~\bibnamefont{Creedon}},
  \bibinfo{author}{\bibfnamefont{S.}~\bibnamefont{Galliou}}, \bibnamefont{and}
  \bibinfo{author}{\bibfnamefont{M.}~\bibnamefont{Tobar}},
  \bibinfo{journal}{Physical Review Letters} \textbf{\bibinfo{volume}{111}},
  \bibinfo{pages}{085502} (\bibinfo{year}{2013}{\natexlab{a}}).

\bibitem[{\citenamefont{Galliou et~al.}(2013)\citenamefont{Galliou, Goryachev,
  Bourquin, Abbe, Aubry, and Tobar}}]{ScRep}
\bibinfo{author}{\bibfnamefont{S.}~\bibnamefont{Galliou}},
  \bibinfo{author}{\bibfnamefont{M.}~\bibnamefont{Goryachev}},
  \bibinfo{author}{\bibfnamefont{R.}~\bibnamefont{Bourquin}},
  \bibinfo{author}{\bibfnamefont{P.}~\bibnamefont{Abbe}},
  \bibinfo{author}{\bibfnamefont{J.}~\bibnamefont{Aubry}}, \bibnamefont{and}
  \bibinfo{author}{\bibfnamefont{M.}~\bibnamefont{Tobar}},
  \bibinfo{journal}{Nature: Scientific Reports} \textbf{\bibinfo{volume}{3}}
  (\bibinfo{year}{2013}).

\bibitem[{\citenamefont{Goryachev and
  Tobar}(2014{\natexlab{a}})}]{Goryachev:2014aa}
\bibinfo{author}{\bibfnamefont{M.}~\bibnamefont{Goryachev}} \bibnamefont{and}
  \bibinfo{author}{\bibfnamefont{M.~E.} \bibnamefont{Tobar}},
  \bibinfo{journal}{New Journal of Physics} \textbf{\bibinfo{volume}{16}},
  \bibinfo{pages}{083007} (\bibinfo{year}{2014}{\natexlab{a}}),
  \urlprefix\url{http://stacks.iop.org/1367-2630/16/i=8/a=083007}.

\bibitem[{\citenamefont{Goryachev and Tobar}(2014{\natexlab{b}})}]{GW}
\bibinfo{author}{\bibfnamefont{M.}~\bibnamefont{Goryachev}} \bibnamefont{and}
  \bibinfo{author}{\bibfnamefont{M.~E.} \bibnamefont{Tobar}},
  \bibinfo{journal}{Physical Review D} \textbf{\bibinfo{volume}{90}},
  \bibinfo{pages}{102005} (\bibinfo{year}{2014}{\natexlab{b}}),
  \urlprefix\url{http://link.aps.org/doi/10.1103/PhysRevD.90.102005}.

\bibitem[{\citenamefont{Pikovski et~al.}(2012)\citenamefont{Pikovski, Vanner,
  Aspelmeyer, Kim, and Brukner}}]{Pikovski}
\bibinfo{author}{\bibfnamefont{I.}~\bibnamefont{Pikovski}},
  \bibinfo{author}{\bibfnamefont{M.~R.} \bibnamefont{Vanner}},
  \bibinfo{author}{\bibfnamefont{M.}~\bibnamefont{Aspelmeyer}},
  \bibinfo{author}{\bibfnamefont{M.~S.} \bibnamefont{Kim}}, \bibnamefont{and}
  \bibinfo{author}{\bibfnamefont{C.}~\bibnamefont{Brukner}},
  \bibinfo{journal}{Nat Phys} \textbf{\bibinfo{volume}{8}},
  \bibinfo{pages}{393} (\bibinfo{year}{2012}),
  \urlprefix\url{http://dx.doi.org/10.1038/nphys2262}.

\bibitem[{\citenamefont{Goryachev et~al.}(2014)\citenamefont{Goryachev, Ivanov,
  van Kann, Galliou, and Tobar}}]{Goryachev:2014ab}
\bibinfo{author}{\bibfnamefont{M.}~\bibnamefont{Goryachev}},
  \bibinfo{author}{\bibfnamefont{E.~N.} \bibnamefont{Ivanov}},
  \bibinfo{author}{\bibfnamefont{F.}~\bibnamefont{van Kann}},
  \bibinfo{author}{\bibfnamefont{S.}~\bibnamefont{Galliou}}, \bibnamefont{and}
  \bibinfo{author}{\bibfnamefont{M.~E.} \bibnamefont{Tobar}},
  \bibinfo{journal}{Applied Physics Letters} \textbf{\bibinfo{volume}{105}},
  (\bibinfo{year}{2014}),
  \urlprefix\url{http://scitation.aip.org/content/aip/journal/apl/105/15/10.1063/1.4898813}.

\bibitem[{\citenamefont{Goryachev et~al.}(January 2012)\citenamefont{Goryachev,
  Galliou, Abbe, Bourgeois, Grop, and Dubois}}]{Goryachev:2012jx}
\bibinfo{author}{\bibfnamefont{M.}~\bibnamefont{Goryachev}},
  \bibinfo{author}{\bibfnamefont{S.}~\bibnamefont{Galliou}},
  \bibinfo{author}{\bibfnamefont{P.}~\bibnamefont{Abbe}},
  \bibinfo{author}{\bibfnamefont{P.}~\bibnamefont{Bourgeois}},
  \bibinfo{author}{\bibfnamefont{S.}~\bibnamefont{Grop}}, \bibnamefont{and}
  \bibinfo{author}{\bibfnamefont{B.}~\bibnamefont{Dubois}},
  \bibinfo{journal}{Ultrasonics, Ferroelectrics and Frequency Control, IEEE
  Transactions on} \textbf{\bibinfo{volume}{59}}, \bibinfo{pages}{21}
  (\bibinfo{year}{January 2012}).

\bibitem[{\citenamefont{Goryachev
  et~al.}(2013{\natexlab{b}})\citenamefont{Goryachev, Galliou, Imbaud, and
  Abb{\'e}}}]{Goryachev:2013ly}
\bibinfo{author}{\bibfnamefont{M.}~\bibnamefont{Goryachev}},
  \bibinfo{author}{\bibfnamefont{S.}~\bibnamefont{Galliou}},
  \bibinfo{author}{\bibfnamefont{J.}~\bibnamefont{Imbaud}}, \bibnamefont{and}
  \bibinfo{author}{\bibfnamefont{P.}~\bibnamefont{Abb{\'e}}},
  \bibinfo{journal}{Cryogenics} \textbf{\bibinfo{volume}{57}},
  \bibinfo{pages}{104} (\bibinfo{year}{2013}{\natexlab{b}}),
  \urlprefix\url{http://www.sciencedirect.com/science/article/pii/S0011227513000532}.

\bibitem[{\citenamefont{Goryachev et~al.}(2011)\citenamefont{Goryachev,
  Galliou, Imbaud, Bourquin, and Abb\'{e}}}]{SunFr}
\bibinfo{author}{\bibfnamefont{M.}~\bibnamefont{Goryachev}},
  \bibinfo{author}{\bibfnamefont{S.}~\bibnamefont{Galliou}},
  \bibinfo{author}{\bibfnamefont{J.}~\bibnamefont{Imbaud}},
  \bibinfo{author}{\bibfnamefont{R.}~\bibnamefont{Bourquin}}, \bibnamefont{and}
  \bibinfo{author}{\bibfnamefont{P.}~\bibnamefont{Abb\'{e}}}, in
  \emph{\bibinfo{booktitle}{Proc EFTF \& IEEE IFCS Joint Meeting}}
  (\bibinfo{address}{San Francisco, USA}, \bibinfo{year}{2011}).

\bibitem[{\citenamefont{M{\"u}ller
  et~al.}(2003{\natexlab{c}})\citenamefont{M{\"u}ller, Herrmann, Saenz, Peters,
  and L{\"a}mmerzahl}}]{ResSME}
\bibinfo{author}{\bibfnamefont{H.}~\bibnamefont{M{\"u}ller}},
  \bibinfo{author}{\bibfnamefont{S.}~\bibnamefont{Herrmann}},
  \bibinfo{author}{\bibfnamefont{A.}~\bibnamefont{Saenz}},
  \bibinfo{author}{\bibfnamefont{A.}~\bibnamefont{Peters}}, \bibnamefont{and}
  \bibinfo{author}{\bibfnamefont{C.}~\bibnamefont{L{\"a}mmerzahl}},
  \bibinfo{journal}{Physical Review D} \textbf{\bibinfo{volume}{68}},
  \bibinfo{pages}{116006} (\bibinfo{year}{2003}{\natexlab{c}}),
  \urlprefix\url{http://link.aps.org/doi/10.1103/PhysRevD.68.116006}.

\bibitem[{\citenamefont{M{\"u}ller et~al.}(2004)\citenamefont{M{\"u}ller,
  Herrmann, Saenz, Peters, and L{\"a}mmerzahl}}]{H2SME}
\bibinfo{author}{\bibfnamefont{H.}~\bibnamefont{M{\"u}ller}},
  \bibinfo{author}{\bibfnamefont{S.}~\bibnamefont{Herrmann}},
  \bibinfo{author}{\bibfnamefont{A.}~\bibnamefont{Saenz}},
  \bibinfo{author}{\bibfnamefont{A.}~\bibnamefont{Peters}}, \bibnamefont{and}
  \bibinfo{author}{\bibfnamefont{C.}~\bibnamefont{L{\"a}mmerzahl}},
  \bibinfo{journal}{Physical Review D} \textbf{\bibinfo{volume}{70}},
  \bibinfo{pages}{076004} (\bibinfo{year}{2004}),
  \urlprefix\url{http://link.aps.org/doi/10.1103/PhysRevD.70.076004}.

\end{thebibliography}

\end{document}